\documentclass[aps, pra, reprint, superscriptaddress, twocolumn, showkeys, amsmath, amssymb,floatfix]{revtex4-2}
\usepackage[english]{babel}
\usepackage{amsmath, amssymb, bbm, mathrsfs, bm, braket, color, graphicx, comment, amsfonts, dsfont}
\usepackage[colorlinks,citecolor=blue,urlcolor=blue]{hyperref}
\usepackage[mathscr]{euscript}
\usepackage{calc}

\usepackage[normalem]{ulem}

\newcommand{\Z}{\mathbb{Z}}

\begin{document}

\title{Topological reflection matrix}
\author{S. Franca}
\affiliation{IFW Dresden and W{\"u}rzburg-Dresden Cluster of Excellence ct.qmat, Helmholtzstr. 20, 01069 Dresden, Germany}

\author{F. Hassler}
\affiliation{JARA-Institute for Quantum Information, RWTH Aachen University, 52056 Aachen, Germany}

\author{I. C. Fulga}
\affiliation{IFW Dresden and W{\"u}rzburg-Dresden Cluster of Excellence ct.qmat, Helmholtzstr. 20, 01069 Dresden, Germany}

\begin{abstract}

While periodically-driven phases offer a unique insight into non-equilibrium topology that is richer than its static counterpart, their experimental realization is often hindered by ubiquitous decoherence effects. 
Recently, we have proposed a decoherence-free approach of realizing these Floquet phases. The central insight is that the reflection matrix, being unitary for a bulk insulator, plays the role of a Floquet time-evolution operator. We have shown that reflection processes off the boundaries of systems supporting higher-order topological phases (HOTPs) simulate non-trivial Floquet phases. 
So far, this method was shown to work for one-dimensional Floquet topological phases protected by local symmetries. 
Here, we extend the range of applicability by studying reflection off three-dimensional HOTPs with corner and hinge modes. 
We show that the reflection processes can simulate both first-order and second-order Floquet phases, protected by a combination of local and spatial symmetries. 
For every phase, we discuss appropriate topological invariants calculated with the nested scattering matrix method. 

\end{abstract}

\maketitle

\section{Introduction} \label{sec:intro}

Symmetries that are present in materials impose constraints on the Hamiltonians that describe them. 
The resulting Hamiltonian can exhibit robust and potentially useful boundary states~\cite{Sarma2015}. 
These states are protected by bulk topological invariants, a connection dubbed the bulk-boundary correspondence~\cite{Hasan2010, Qi2011}. 
Theoretically, the possible invariants depend on the symmetries and the dimensionality of the system. 
They are given by integers and may change their value only when the bulk gap closes. 
The type of the topological invariant~\cite{Thouless1982,Kane2005} ($\Z_2$ or $\Z$) determines whether the system supports only one topological phase ($\Z_2$), or there are more topological phases, distinguished by a different number of symmetry protected boundary modes ($\Z$). 

In a static $D$-dimensional system described by a Hermitian Hamiltonian operator, the presence of particle-hole ($\cal{P}$) and/or chiral ($\cal{C}$) symmetries renders the energy spectrum symmetric around zero energy. 
As a result, robust $(D-1)$-dimensional boundary states can only appear at zero energy with $E=0$. 
Since these are local symmetries, spectrally-isolated boundary modes will be present simultaneously at all surfaces of the system. 
The corresponding system is then in a strong topological phase (STP). 
This by far thus not exhaust all possible topological states. 
Adding spatial symmetries (that act non-locally and relate different sites of the system), a topological crystalline phase (TCP)~\cite{Fu2011, Hsieh2012} can be realized with an adequate crystalline symmetry, or a weak topological phase (WTP) with a suitable translation symmetry~\citep{Fu2007}. 
Another possibility that enriches the classification are $n^{\rm th}$-order ($n>1$) topological phases, with $(D-n)$-dimensional gapless states~\cite{Benalcazar2017, Schindler2018}. 
Here, we dub all $n>1$ topological systems higher-order topological phases (HOTPs).
From the listed types of topological phases, STPs are the most stable in the presence of disorder because they only rely on the presence of local symmetries~\cite{Fu2007, Ringel2012}.

Topological considerations are not restricted to equilibrium systems. 
In particular, periodically-driven (or Floquet) phases have been studied extensively~\cite{Kitagawa2011, Gomez-Leon2013, Kitagawa2010, Leykam2016}. 
Their topological properties are determined from the knowledge of the unitary Floquet operator $\mathcal{F}$ that is the time-evolution operator over the time period $T$. 
In this case, the relevant spectrum is the $2\pi/T$ periodic quasienergy spectrum constructed from the eigenphases of the Floquet operator; in the following, we set $T=1$ for convenience. Note that Floquet phases are classified using unitary operators while conventional static systems are classified using Hermitian operators.
Similar to static systems, symmetries impose constraints on the quasienergy spectrum. 
There are two eigenphases ($0$ and $\pi$) left invariant under the action of $\cal{P}$ and $\cal{C}$. Here, different from static systems, topological boundary modes can occur either at one of these eigenphases or at both. 
In the latter case, the system is in an anomalous Floquet phase characterized by topological indices beyond those possible in Hermitian systems~\cite{Kitagawa2010, Leykam2016, Zhou2016, Quelle2017, Jiang2011, Rudner2013}. 
Note that experimental realizations of Floquet topological phases are much more demanding compared to their static counterparts. 
The main problems, decoherence and heating, are inherently present in most of the driving processes~\cite{Groh2016, Rieder2018, Sieberer2018, Lazarides2014,DAlessio2014}. 

Despite differences between static and Floquet systems, scattering theory provides a unified framework for their topological characterization~\cite{Akhmerov2011, Fulga2011, Fulga2012, Fulga2016}. 
This theory can also be used as a tool to perform dimensional reduction in a two-terminal geometry, as the reflection matrix of a $D$-dimensional system effectively describes a $(D-1)$-dimensional system \cite{Fulga2012, Geier2018, Franca2021}. 
With this tool, Hermitian systems exhibiting topological phases of the same order $n$ but belonging to different symmetry classes~\citep{Altland1997} have been related following the Bott periodicity~\cite{Fulga2012, Geier2018}. 
In our previous work~\cite{Franca2021}, we have introduced an alternative dimensional reduction procedure that works for topologically nontrivial systems with gapped bulk and gapless boundary states that do not conduct between the leads. 
This requirement is accommodated by many $D$-dimensional systems exhibiting $n^{\rm th}$-order topological phases ($1<n\leq D$). 
The dimensional reduction maps the static system onto its unitary reflection matrix. 
The latter describes a $(D-1)$-dimensional Floquet system in a $(n-1)^{\rm th}$-order topological phase, but in the same symmetry class as the static system. 

From a practical point of view, this dimensional reduction procedure implies that we can use Hamiltonian systems in higher-order topological phases to simulate lower-dimensional Floquet topological phases~\cite{Franca2021}. 
Such an approach of simulating Floquet systems is advantageous to implementing the phase in a driven(-dissipative) system as it eliminates the need for an external driving field, as well as decoherence related problems caused by the noise in the driving field. Indeed, noise-induced decoherence in single-particle driven systems is by now a well-recognized problem, and a large number of publications address this issue, both theoretically and experimentally \cite{Ammann1998, Klappauf1998, Steck2000, dArcy2001, Oskay2003, Sadgrove2004, White2014, Bitter2016, Bitter2017, Sarkar2017, Jrg2017, Rieder2018, Sieberer2018, ade2019, Bomantara2020, Wintersperger2020, Timms2020, Ravindranath2021}.

Experimental verification of our results is based on two prerequisites: (1) the experimental realization of HOTPs~\cite{Noh2018, Xie2018, Xie2019, Serra-Garcia2018, Xue2018, Xue2019, Zhang2019, Zhang2019a, Xue2019a, Ni2019, Chen2019, Zhang2020, Peterson2018, Imhof2018, Kempkes2019}, and (2) the measurement of the eigenphases of the reflection matrix~\cite{Hu2015, Wang2017, Kempkes2019, Laforge2019}. 
So far, acoustic~\cite{Xue2018, Xue2019, Zhang2019, Zhang2019a, Xue2019a, Ni2019, Chen2019, Zhang2020}, photonic~\cite{Noh2018, Xie2018, Xie2019}, phononic~\cite{Serra-Garcia2018}, microwave~\cite{Peterson2018}, topoelectric~\cite{Imhof2018}, and condensed-matter~\cite{Kempkes2019} platforms were successfully adapted to realize HOTPs. 
For some of these systems, the reflection matrix eigenvalues can be directly visualized~\cite{Kempkes2019, Laforge2019}, while for others they can be inferred from interferometric techniques~\cite{Hu2015, Wang2017}. 

Motivated by the possibility to simulate periodically-driven phases as the reflection properties of static systems, we study different realizations of HOTPs that, once mapped to lower-dimensional counterparts, realize various kinds of Floquet phases. 
In particular, we first consider a prototype of the three-dimensional (3D) system with zero-energy corner modes, a Benalcazar-Bernevig-Hughes (BBH) model~\cite{Benalcazar2017}. 
By modifying this system in one spatial direction, it can also realize a second-order topological phase with hinge modes. 
We consider cases when these hinge states are protected by either translation or point group symmetries, in addition to the local symmetries. 
This presents an advancement compared to Ref.~\cite{Franca2021}, which studied only scattering regions with gapless corner states that are robust to any spatial symmetry breaking. We thus show that reflection matrices can simulate a wider range of Floquet topological phases than realized in the previous work e.g., weak or crystalline Floquet topological phases.
Given the fact that the classification of Floquet phases only relies on the matrices being unitary, we will call them \emph{unitary phases} in the following.
 
The rest of the paper is organized as follows. 
In Sec.~\ref{sec:methods}, we discuss details of the scattering setup and present the symmetry constraints obeyed by the scattering/reflection matrix. In Sec.~\ref{sec:BBH}, we study reflection matrices of different 3D systems and show they realize first- and second-order unitary topological phases. We characterize these unitary systems in Sec.~\ref{sec:char} using a nested scattering matrix procedure. Finally, in Sec.~\ref{sec:conclusion}, we discuss our results and outline directions for future research.  

\section{Methods} \label{sec:methods}

In this Section, we first review the scattering setup used to simulate unitary topological phases before we discuss the role of local symmetries. 

\subsection{Scattering setup} \label{subsec:Scatt_setup} 

\begin{figure}[tb]
\includegraphics[width=0.9\columnwidth]{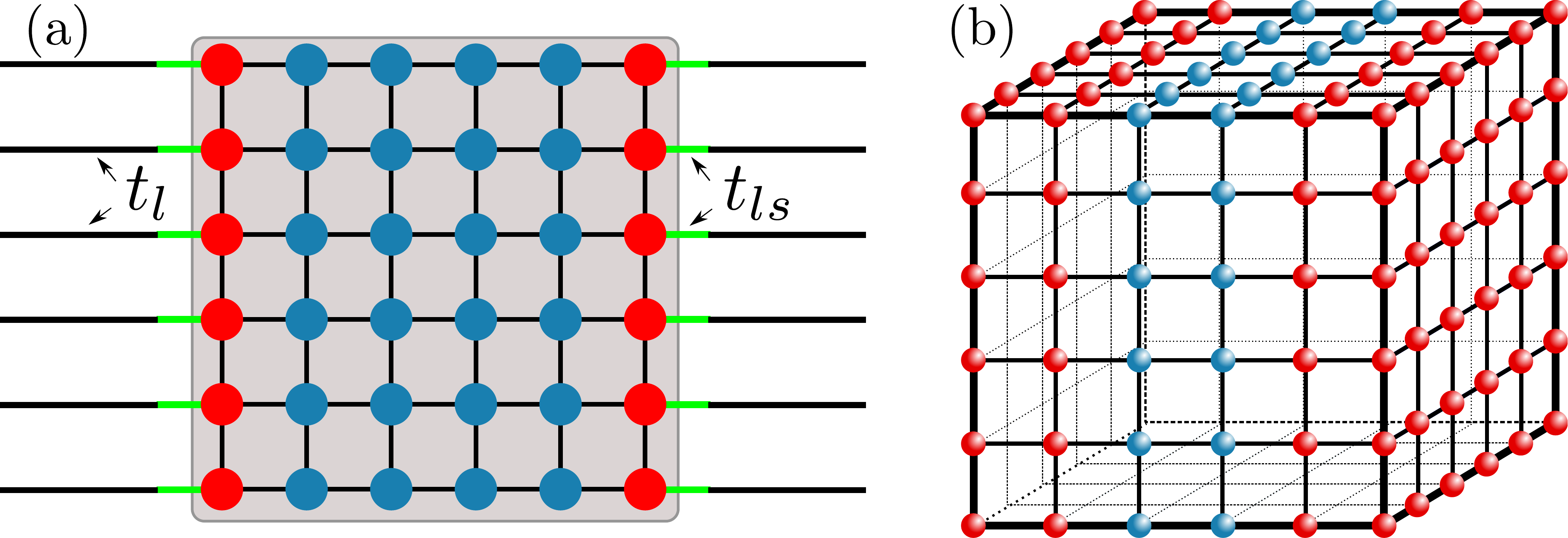}
 \caption{(a) Sketch of the two-lead geometry for a two-dimensional scattering region, represented by a gray background. 
 Here, the sites of the system are represented by blue dots while the red dots denote those sites to which leads are attached. 
 This setup has $N_d = 1$ (see main text for its definition), meaning that each lead is only attached to the first layer of sites. The leads are described by idealized waveguides, represented with black lines. They are connected to the system via weak links, colored in green. 
 (b)  3D scattering system with $N_d = 2$, indicating that each lead is attached to two layers of the system. 
 For clarity, we have plotted only sites on the visible surfaces of the 3D system and have omitted plotting the waveguides.}
\label{fig:setup}
\end{figure}

The scattering setup consists of a static system that supports a HOTP, and two translationally invariant leads attached to opposite surfaces of this system, see Fig.~\ref{fig:setup}. 
Unless otherwise specified, we orient the leads along the \textit{x} direction, and denote them as the left and right leads. 
The scattering matrix relates incoming to outgoing lead modes via $\psi_{\rm out} = S \,\psi_{\rm in}$. In a two-terminal geometry assumes the general form 
\begin{eqnarray}\label{eq:scat_matrix}
S=\begin{pmatrix}r&t'\\t&r'\end{pmatrix};
\end{eqnarray}
here, $r$ and $r'$ are the reflection matrices of the left and right leads, respectively, while $t$ and $t'$ are the transmission matrices.  In the following, we consider transport at $E=0$, which is the special point for the particle-hole and the chiral symmetry.

We construct the left and right leads out of arrays of decoupled and translationally invariant waveguides (chains). 
The Hamiltonian of the $i$-th chain reads $\mathcal{H}_i = \sum_{k_x} t_l \sin{k_x} c_{k_x}^{\dagger} c_{k_x}$, where $ c_{k_x}^{\dagger}(c_{k_x})$ is the electron creation (annihilation) operator and $t_l > 0$ is the hopping amplitude along each chain. Here, the index $i$ denotes the transversal (spatial) position of the waveguide.
We assume that each waveguide probes $N_d$ degrees of freedom located on or near the surface of the system. 
In particular, in the following we are interested in staggered systems such that $N_d=1$ probes half of the unit cell and $N_d=2$ the full unit cell. 
As a result, the (effective) number of incoming (outgoing) modes per waveguide equals $N_d$, see Fig.~\ref{fig:setup}. 
We define the interface region as the set of all sites to which a chain is attached. 
The reflection process at transversal site $i$ of the left interface region is described by the $N_d\times N_d$ matrix $r_{ii}$, which is a diagonal sub-block of $r$; in the following, we denote this $N_d \times N_d$ scattering matrix as an \emph{element} of the scattering matrix. 
Similarly, there are the reflection elements $r_{ji}$ that describe an incoming mode at site $i$ that is reflected back into the site $j$. 
Due to the fact that we are considering systems which are insulating in the bulk and at surfaces, we expect $r_{ji}$ to decay as a function of distance between $i$ and $j$.
This means that $r$ is endowed with a spatial structure; the incoming and outgoing modes can be labeled by the real space positions $i$ along the transversal direction. 
Since the leads are attached to the surfaces of a $D$-dimensional system, the reflection matrix probes only the surface and the index $i$ naturally has dimension $(D-1)$.  
This special way of designing the lead modes provides access to the transversal dimension of the system. It simplifies the discussion and allows us to draw conclusions about the spatial position of the relevant modes. 
Further, this lead construction is directly implementable in meta-materials, which so far have been the main platforms realizing HOTPs, as we have mentioned in the introduction. 
When considering topological electronic circuits~\cite{Imhof2018}, for instance, such a lead can be created by connecting one electric cable to each of the circuit nodes representing the boundary of the system.
Note, however, that this specific construction is irrelevant for the topological classification (being basis independent) and thus, experimentally, also different mode structures are allowed.

In our previous work~\cite{Franca2021}, we have shown that the reflection process from a boundary of a static 2D BBH system can simulate four different one-dimensional (1D) unitary phases (via the natural identification $\mathcal{F} \Leftrightarrow r$). 
The kind of unitary phase that is realized depends on the parameters of the BBH model and $N_d$. 
If the leads are attached to the outermost sites of the scattering region ($N_d = 1$), the reflection matrix simulates the trivial phase or topological phases with either $0$ modes or $\pi$ modes. 
The analogue of an anomalous unitary phase is obtained when the reflection matrix is `thicker' and probes the full unit cell, consisting of two layers of sites, i.e. $N_d =2$. 
In Fig.~\ref{fig:setup}(a) and (b) we show a setup with $N_d = 1$ and $N_d =2$, respectively.

We note in passing that, in experiments, it is important that the scattering setup is designed such that the leads only weakly perturb the system. 
This requirement is realized if the coupling between leads and the system $t_{ls}$ is weak, i.e., such that the level broadening $\Gamma \simeq t_{ls}^2/t_l$ it is smaller than the energy gap of the system. 
In our numerical simulations performed with the Kwant code~\cite{Groth2014}, the system is always in its ground state. 
We have thus taken $t_{ls} = t_l$ to ensure a good visibility of topological features.

\subsection{Symmetry constraints}~\label{sec:symm}

Simulating Floquet physics using the unitary reflection matrix $r$ requires that $r$ and the Floquet operator $\mathcal{F}$ obey the same symmetry constrains. 
We first discuss the case of the onsite symmetries: $\mathcal{T}$, $\mathcal{P}$, and $\mathcal{C}$, with $\mathcal{T}$ the time-reversal symmetry.
They can be written as $\mathcal{T} = U_{\mathcal{T}} \mathcal{K}$, $\mathcal{P} = U_{\mathcal{P}} \mathcal{K}$, and $\mathcal{C} = U_{\mathcal{C}}$, where $U_{\mathcal{T}}, U_{\mathcal{P}}, U_{\mathcal{C}}$ are unitary operators and $\mathcal{K}$ denotes complex conjugation. 
To determine the restrictions imposed on $S$, $r$, and $\mathcal{F}$ by these symmetries, it is crucial to know how they act on the Hamiltonian $H$~\citep{Altland1997}.

Since the scattering matrix $S$ is calculated from the Schr{\"o}dinger equation describing the full, system-plus-lead problem~\citep{Fulga2012, Franca2021}, it has been shown that~\footnote{These relations can be obtained by simple algebraic manipulations of Eqs.~(A11a-A11c) of Ref.~\cite{Fulga2012}, using $V_{\mathcal{P},\mathcal{C}} = U_{\mathcal{P},\mathcal{C}}$.}
\begin{equation}\label{eq:S_symm}
 U_{\mathcal{T}} S^* U_{\mathcal{T}}^{\dagger} = S^{\dagger}, \;
 U_{\mathcal{P}} S^* U_{\mathcal{P}}^{\dagger} = S, \;
 U_{\mathcal{C}} S^{\dagger} U_{\mathcal{C}}^{\dagger} = S.
\end{equation}

On the other hand, the Floquet operator constraints follow from the definition of this operator $\mathcal{F} = \overline{\exp}{(-i \int_0^{1} H(t) dt)}$, where $\overline{\exp}$ denotes the time-ordered exponential~\cite{Kitagawa2010,Fulga2016}.
This gives
\begin{equation}\label{eq:F_symm}
 U_{\mathcal{T}} \mathcal{F}^* U_\mathcal{T}^{\dagger} = \mathcal{F}^{\dagger}, \;
 U_{\mathcal{P}} \mathcal{F}^* U_\mathcal{P}^{\dagger} = \mathcal{F}, \;
 U_{\mathcal{C}} \mathcal{F}^{\dagger} U_{\mathcal{C}}^{\dagger} = \mathcal{F}.
\end{equation} 

We see that both $S$ and ${\cal F}$ behave identically under the action of local symmetries. 
If the scattering region does not conduct between the leads, the only nonzero parts of $S$ are its reflection matrices. 
Hence, $r$ inherits the symmetry and the unitarity from $S$ and can be directly interpreted as $\mathcal{F}$ (having the correct symmetries).

Spatial symmetries $\mathcal{U}$ impose additional restrictions on incoming/outgoing modes. 
We consider the ones that act in a plane parallel to the system-lead interface, which is spanned by a momentum $\bm{k}_{\parallel}$. The Schr{\"o}dinger equation provides the constraint
\begin{align}\label{eq:spatial_symm}
\begin{split}
\mathcal{U} S (\bm{k}_{\parallel}) \mathcal{U}^{\dagger} =  S ( {\cal R} \bm{k}_{\parallel}),
\end{split}
\end{align}
where ${\cal R}\bm{k}_{\parallel}$ is the transformed momentum vector due to the action of the spatial symmetry.

\section{Results} \label{sec:BBH}

In this Section, we extend the results of Ref.~\cite{Franca2021}, focusing on 3D Hamiltonian systems in second- and third-order topological phases. 
We show that unitary reflection matrices can be interpreted as Floquet operators that realize first- and second-order topological phases. 
To make the rest of this section more accessible, we begin by highlighting our results and the types of scattering problem we consider, before showing the details of the calculations.

\begin{figure}[tb]
\includegraphics[width=1\columnwidth]{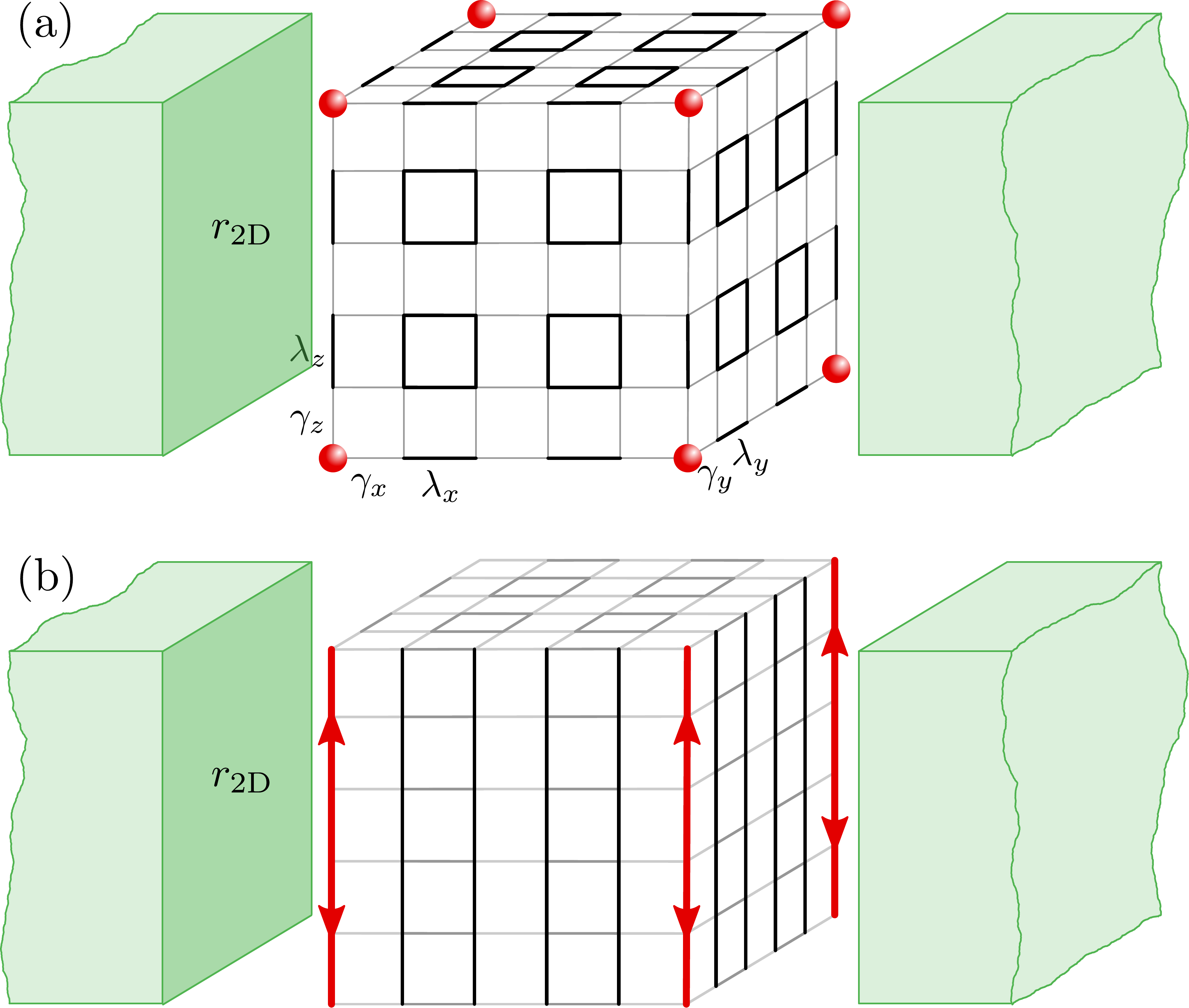}
 \caption{Sketch of 3D finite-sized HOTPs. The unequal hoppings $\gamma_{\alpha}$ and $\lambda_{\alpha}$ ($\alpha = x,y,z$) lead to staggering indicated by the thickness of the lines. The topological zero modes are shown in red. In panel (a), $|\gamma_{\alpha}| < |\lambda_{\alpha}|$ and the system exhibits corner states at $E=0$. In panel (b), $|\gamma_{x,y}| < |\lambda_{x,y}|$ but $|\gamma_{z}| = |\lambda_{z}|$, so the system hosts counter-propagating modes on its vertical hinges.}
\label{fig:sys3d}
\end{figure}

In Sec.~\ref{sec:corner}, we consider a 3D HOTP with eight gapless corner modes, as shown in Fig.~\ref{fig:sys3d}(a). 
The system is an insulator, so there is no transmission between the two leads attached to left and right surfaces lying in the \textit{yz} plane (and hence called \textit{yz} surfaces). 
As a result, the reflection matrix that relates incoming and outgoing modes of the lead is unitary. 
Due to the way we constructed our leads (see Sec.~\ref{subsec:Scatt_setup}), it is possible to label every incoming plane wave according to its position in the \textit{yz} plane. 
If it impinges on a zero-energy state, it will be reflected with a $\pi$-phase shift due to a resonant scattering~\citep{Akhmerov2011}. 
Thus, incoming and outgoing plane waves at the corners of the lead differ by a $\pi$-phase shift. 
The plane waves reflecting from the other sites on the surface of this system will experience position-dependent phase shifts that assume non-quantized values and form bands~\cite{Franca2021}. 
In total, there are four resonant scatterings per lead, corresponding to four $\pi$ modes in the eigenphase spectrum of $r$. 
Every $\pi$ mode is pinned to a corner of the lead, and thus the reflection matrix $r$ is a 2D Floquet operator in a second-order unitary phase with $\pi$ modes at the corners~\cite{Rodriguez-Vega2019, Bomentara2019}.

The remainder of Sec.~\ref{sec:BBH} deals with two kinds of 3D systems with gapless and dispersing hinge modes. These modes enable conduction along the hinges, and therefore the reflection matrices of these 3D systems are not unitary for all scattering setups, as it was the case for a scattering region with corner modes. However, an example of a scattering setup that does produce a unitary reflection matrix is shown in Fig.~\ref{fig:sys3d}(b). Here, we see that leads are oriented perpendicular to the direction of the hinge modes, such that they detect no conduction. 

The lead detects hinge modes of the 3D HOTP that translate into dispersing boundary modes in the eigenphase spectrum of the reflection matrix. Whenever the hinge modes of the 3D HOTP cross zero energy, the boundary modes of $r$ will show $\pi$ crossings in the eigenphase spectrum. 
If the Hamiltonian hinge modes are protected by translation (or some crystalline) symmetry in addition to a local symmetry, then $r$ implements a 2D unitary WTP (or TCP).

\subsection{3D system with corner modes}\label{sec:corner}

As a model of a 3D system with corner modes, we consider the 3D BBH model. This is a tight-binding model of spinless electrons on a cubic lattice, as represented in Fig.~\ref{fig:sys3d}(a).
The hoppings between nearest-neighbor sites are staggered in all three directions, such that the resulting cubic unit cell has eight sites in total. 
Some of these hoppings have negative values, in order to ensure a  flux of $\pi$  threading each face of the cube.

In momentum space with wavevector ${\bm k} = (k_x,k_y, k_z)$, the Fourier transform of the real space Hamiltonian shown in Fig.~\ref{fig:sys3d}(a) reads~\cite{Benalcazar2017}
\begin{align}\label{eq:BBHHam3D}
\begin{split}
h_{\rm 3D}(\bm{k}) {} &= (\gamma_x  + \lambda_x \cos{k_x}) \eta_z \tau_0 \sigma_x + \lambda_x \sin{k_x} \eta_z \tau_0 \sigma_y \\
& + (\gamma_y  + \lambda_y \cos{k_y}) \eta_x \tau_0 \sigma_0 + \lambda_y \sin{k_y} \eta_y \tau_z \sigma_0 \\
& + (\gamma_z  + \lambda_z \cos{k_z}) \eta_y \tau_y \sigma_0  - 
\lambda_z \sin{k_z} \eta_y \tau_x \sigma_0,
\end{split}
\end{align}
where $\gamma_{\alpha}$ and $\lambda_{\alpha}$ are the intracell and intercell hoppings in the $\alpha$ ($\alpha = x,y,z$) direction. 
The Pauli matrices $\sigma$, $\tau$, and $\eta$ denote the sublattice degrees of freedom, corresponding to the 8 sites per unit cell.

\begin{figure}[b]
\includegraphics[width=0.9\columnwidth]{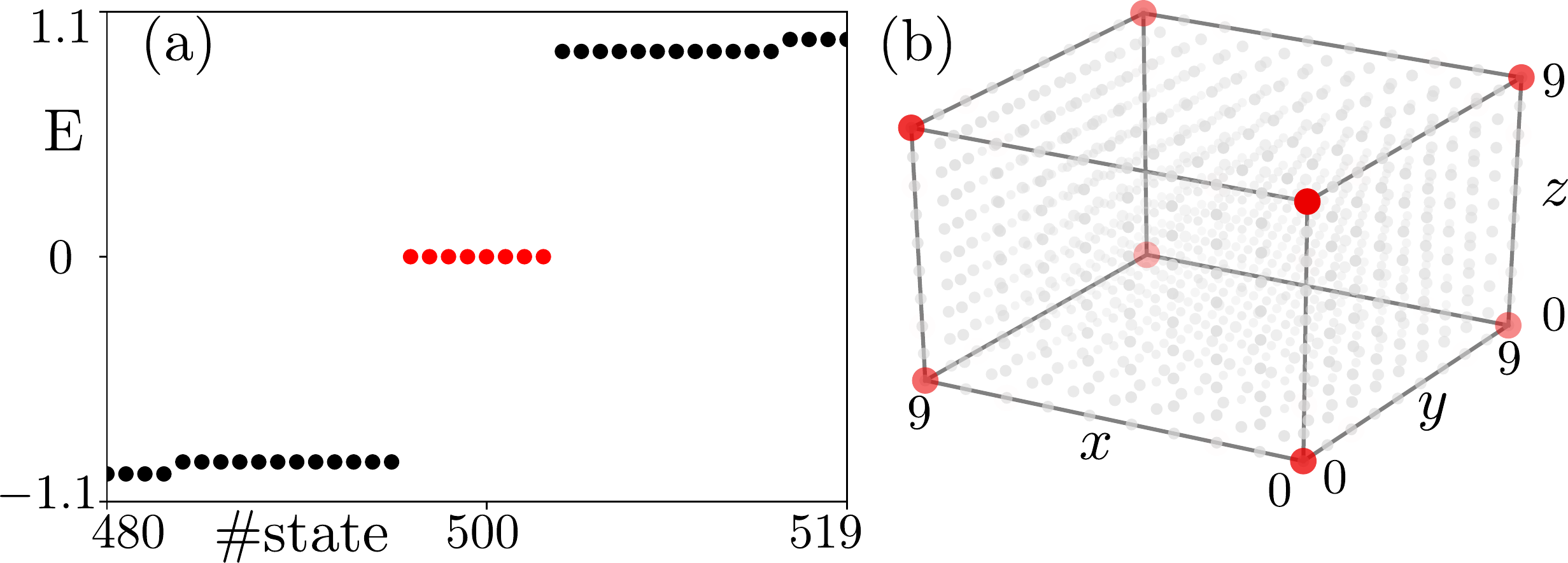}
 \caption{In panel (a), we plot the low-energy spectrum of a 3D system in the HOTP, with eight zero modes colored in red. 
 Panel (b) shows the probability distribution of these midgap modes. 
 We assume the system is cubic, with $L=10$ sites in each direction with $\gamma_{\alpha} = 0.1$ and $\lambda_{\alpha} = 1$.  }
\label{fig:3DHOTI}
\end{figure}  

The Hamiltonian in Eq.~\eqref{eq:BBHHam3D} obeys the symmetries $\mathcal{T} = \mathcal{K}$, $\mathcal{P} = \eta_z \tau_0 \sigma_z \mathcal{K}$ and $\mathcal{C} = \eta_z \tau_0 \sigma_z$, i.e., $[ h, \mathcal{T} ] = \{h, \mathcal{P}\} = \{h, \mathcal{C}\} =0$.
On a finite system with $L^3$ sites, it supports eight zero modes provided that $|\gamma_{\alpha}|<|\lambda_{\alpha}|$; see Fig.~\ref{fig:3DHOTI}(a). These zero modes are corner states, as can be deduced from their probability distribution plotted in Fig.~\ref{fig:3DHOTI}(b). The existence of corner modes and their number is related to the nontrivial value of the topological index~\cite{Benalcazar2017}. The presence of $\mathcal{P}$ ($\mathcal{C}$) symmetry forces this index to be of the $\Z_2$ ($\Z$) type. However, having a single zero-energy state per corner makes the value taken by the particle-hole invariant equal to the one taken by the chiral invariant. For this reason, one of these two symmetries  can be treated as redundant. Thus, in the following, we discuss our results only in terms of the particle-hole symmetry.

The topology of this system~\cite{Benalcazar2017} is related to the dimerization patterns (topological or trivial) of the hinges, akin to the topology of the Su-Schrieffer-Heeger (SSH) model~\cite{Su1979} (see Fig.~\ref{fig:sys3d}). 
If the hopping strengths are independent of direction, $\gamma_{\alpha} = \gamma$ and $\lambda_{\alpha} =\lambda$, the bulk gap closes at $|\gamma| = |\lambda|$, marking a topological transition between a HOTP ($|\gamma| < |\lambda|$) and a trivial phase ($|\gamma| > |\lambda|$) . 
On the other hand, if the hoppings become direction dependent, then the phase transition between a HOTP and a phase without zero-energy corner modes may then occur via a boundary (either hinge or surface) gap closing.
For example, setting $|\gamma_z| = |\lambda_z|$ while keeping $|\gamma_{x,y}| < |\lambda_{x,y}|$ creates two counter-propagating modes along every hinge in the \textit{z} direction, shown in Fig.~\ref{fig:sys3d}(b).

\begin{figure}[tb]
\includegraphics[width=1.0\columnwidth]{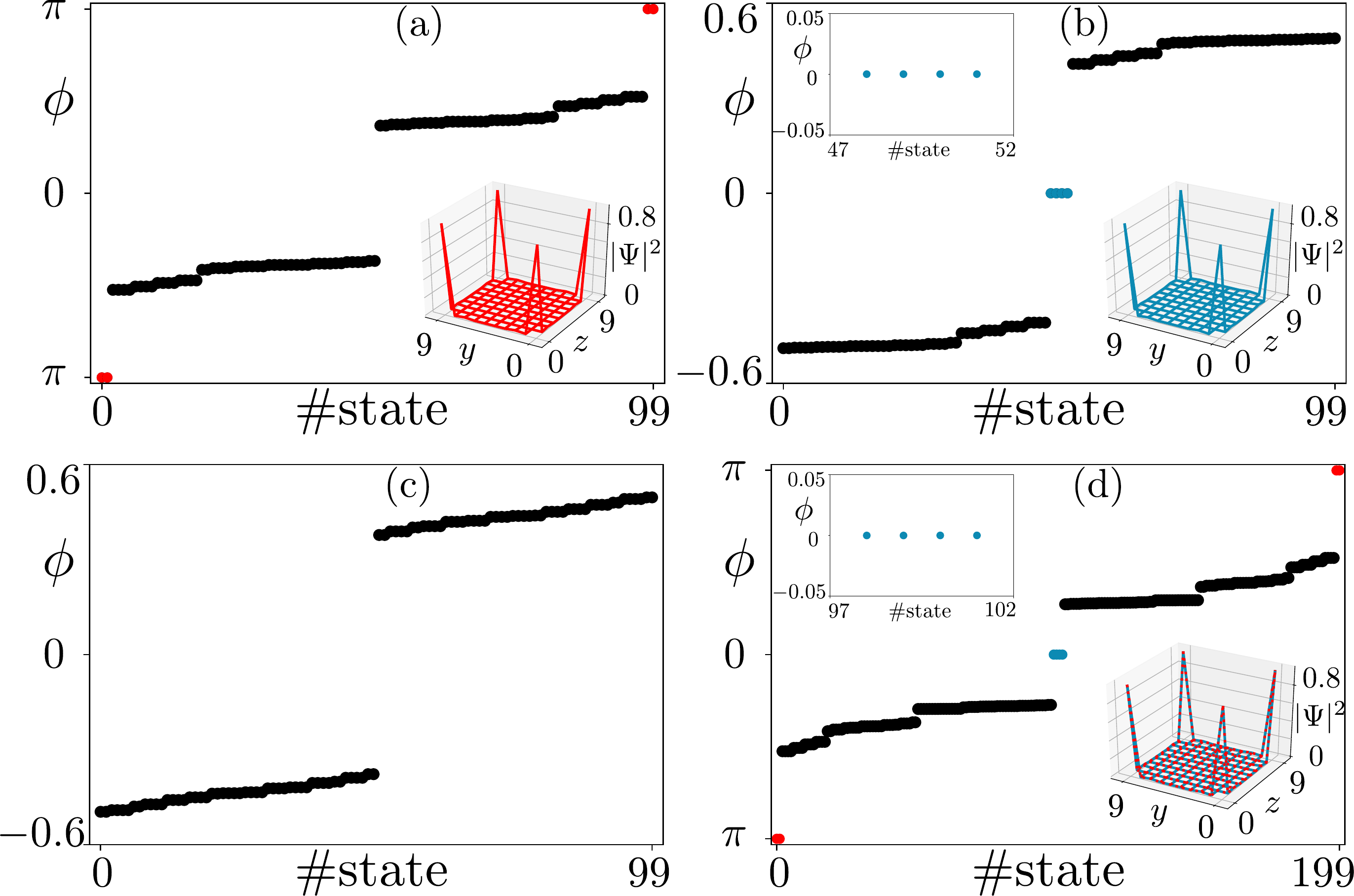}
 \caption{The eigenphase spectra of $r_{\text{2D}}$ are plotted for different dimerizations and interface thicknesses for $L=10$. 
   In the first three panels $N_d = 1$ indicating that each of the two leads is attached only to the first layer of boundary sites. As a result, the matrix $r_{\text{2D}}$ contains $100$ entries in total. 
 In panel (a), we choose $\gamma_{\alpha} = 0.1$ and $\lambda_{\alpha} = 1$. 
 In panel (b), $\gamma_x = 2.1$, $\gamma_{y,z} = 0.1$, $\lambda_{\alpha} =1$. 
 In panel (c),  $\gamma_{\alpha} = 2.1$, $\lambda_{\alpha} = 1$. 
 Panel (d) corresponds to panel (a) except for $N_d = 2$ such that $r_{\text{2D}}$ contains $200$ entries in total. 
 Red (blue) dots denote the $\pi$ (0) modes. 
 The probability distributions of topologically protected modes are given in the insets. }
\label{fig:3DHOTI_r}
\end{figure}

We construct a unitary system by attaching two leads probing opposite \textit{yz} surfaces of the 3D system. 
We set the interface region to be only one layer (half of a unit-cell) of surface sites thick, i.e., $N_d = 1$. In the following, we study the reflection matrix $r_{\text{2D}}$ of the left lead (the result $r'_\text{2D}$ for the right lead is equivalent).

For a system with $|\gamma_{\alpha}|<|\lambda_{\alpha}|$, the eigenphase spectrum of $r_{\text{2D}}$ is plotted in Fig.~\ref{fig:3DHOTI_r}(a). 
There are four modes at eigenphase $\phi=\pi$ shown in red that are pinned to the corners of the 2D lead. 
We can define a quantity $\xi^{0,\pi}  = (-1)^{N_{\rm modes}^{0,\pi}}$ that counts the parity of the number of isolated modes (denoted as $N^{\phi}_{\rm modes}$) per corner at $\phi = 0, \pi$. For Fig.~\ref{fig:3DHOTI_r}(a), we obtain $\xi^{\pi} = -1$ and $\xi^{0} = 1$. In the following, we use this quantity as a topological index. In Sec.~\ref{sec:char}, we will see how is it related to the invariant calculated within the scattering theory. 

Next, we consider a system that has a trivial (intracell) dimerization along the $x$ edges ($|\gamma_x|> |\lambda_x|$) while $|\gamma_{y,z}| < |\lambda_{y,z}|$
along the other edges. Note that the Hamiltonian system does not support zero energy modes in this case.
The eigenphase spectrum of its reflection matrix $r_{\text{2D}}$ is plotted in Fig.~\ref{fig:3DHOTI_r}(b). 
There are four $0$ modes localized at the corners of the lead indicating that the reflection matrix detects the nontrivial topology of the \textit{yz} surface~\cite{Franca2021}. This setup therefore allows $r_{\text{2D}}$ to simulate a second-order Floquet phase with $0$ modes~\cite{Rodriguez-Vega2019, Bomentara2019}. A single $0$ mode per corner of a 2D system leads to $\xi^{\pi} = 1$ and $\xi^{0} = -1$.

If the 3D scattering region has a trivial dimerization along all edges $(|\gamma_{\alpha}| > |\lambda_{\alpha}|)$, the eigenphases are not quantized to the values $0$ and $\pi$ anymore. In fact, all eigenvalues of $r_{\text{2D}}$ form complex-conjugate pairs, as seen in Fig.~\ref{fig:3DHOTI_r}(c), such that the reflection matrix is topologically trivial and $\xi^0 = \xi^\pi = 1$.

Finally, a 2D anomalous second-order unitary phase with corner modes occurring simultaneously at $\phi = 0$ and $\phi = \pi$~\cite{Bomentara2019, Rodriguez-Vega2019} can be realized by coupling a lead to the full unit cell (with $N_d=2$) of a HOTP ($|\gamma| < |\lambda|$). In Fig.~\ref{fig:3DHOTI_r}(d),
we plot the $\phi$ spectrum of such $r_{\text{2D}}$. 
Since there are four corner modes at $\phi=0$ and at $\phi = \pi$, the indices read $\xi^0 = -1$ and $\xi^{\pi} = -1$.  
Thus choosing different parameters in \eqref{eq:BBHHam3D}, it is possible to realize four distinct $2D$ second-order Floquet phases. In fact, these four phases are the only possible ones in symmetry class D when the four-fold rotation symmetry is enforced as it is the case for all our unitary 2D systems. This symmetry constraint follows from Eq.~\eqref{eq:spatial_symm} and the fact that we considered 3D scattering regions with $\mathcal{C}^x_4$ symmetry resulting from setting $\gamma_x = \gamma_y$ and $\lambda_x = \lambda_y$. When present, the four-fold rotation symmetry implies that topological corner states occur at the same eigenphase(s) for all the corners of a square shaped system.

The presence or absence of an anomalous topological phase for $N_d=2$ versus $N_d=1$ is related to the parity of the number of orbitals per site in the reflection matrix.
To see this, consider the maximally dimerized limit, $|\gamma|\to 0$ in Fig.~\ref{fig:sys3d}(a), in which zero-energy corner modes are localized on a site and decoupled from the rest of the system.
The presence of these states results in the formation of $\phi=\pi$ modes in the spectrum of the reflection matrix, due to resonant reflection.
However, since $r_{\text{2D}}$ is real, its eigenvalues can only be real or come in complex-conjugate pairs.
As such, for $N_d=2$, the decoupled corner state of the 3D system must produce both a $\pi$-mode and a $0$-mode in $r_{\text{2D}}$ in order to satisfy the even parity of reflection matrix orbitals.
This leads to an anomalous phase, as shown in Fig.~\ref{fig:3DHOTI_r}(d).
In contrast, when $N_d=1$, $\pi$-modes and $0$-modes cannot form simultaneously due to the odd parity of orbitals [see Fig.~\ref{fig:3DHOTI_r}(a-c)].
The above argument generalizes to thicker interface regions. 
Thus, we expect the nontrivial phases of $r_{\text{2D}}$ to be anomalous when $N_d$ is even and not anomalous when $N_d$ is odd.

Note that in Appendix~\ref{app:R_Green}, we explore the topological phase transitions between unitary phases with and without topological states in more detail. 
There, we also verify our numerical results using analytical reflection matrices calculated with the boundary Green's function technique~\cite{Peng2017}. 
In Appendix \ref{app:lead_coupling}, we examine the effect of a transversal coupling on the waveguide modes, such that the leads are no longer formed out of decoupled chains.

\subsection{3D system with hinge modes protected by a translation symmetry }\label{sec:hinge}

In the following, we investigate reflection matrices of a 3D second-order topological phase with gapless hinge states.
A weak HOTP with hinge modes can be realized by coupling a stack of 2D BBH systems~\cite{Benalcazar2017} with a nearest-neighbor hopping $\gamma_z$ along the $z$ direction, as shown in Fig.~\ref{fig:sys3d}(b). 
The momentum space Hamiltonian of this 3D system reads
\begin{align}\label{eq:BBHHam3D_WTP}
\begin{split}
 h_{\text{WTP}}(\bm{k}) {} = h_{\text{2D}}(\bm{k}) + \gamma_z \sin{k_z} \eta_0 \sigma_0,
\end{split}
\end{align} 
where
\begin{align*}
\begin{split}
 h_{\text{2D}}(\bm{k}) {} & = (\gamma_x  + \lambda_x \cos{k_x}) \eta_z \sigma_x + \lambda_x \sin{k_x} \eta_z  \sigma_y \\
& + (\gamma_y  + \lambda_y \cos{k_y}) \eta_x \sigma_0 + \lambda_y \sin{k_y} \eta_y  \sigma_0
\end{split}
\end{align*}
is the Hamiltonian of a 2D BBH system. 

The Hamiltonian Eq.~\eqref{eq:BBHHam3D_WTP} is particle-hole and chiral symmetric. The latter symmetry can be treated as redundant as explained in Sec.~\ref{sec:corner}. We therefore focus on the $\mathcal{P}$ symmetry related topological protection in the following. 
  
\begin{figure}[tb]
\includegraphics[width=\columnwidth]{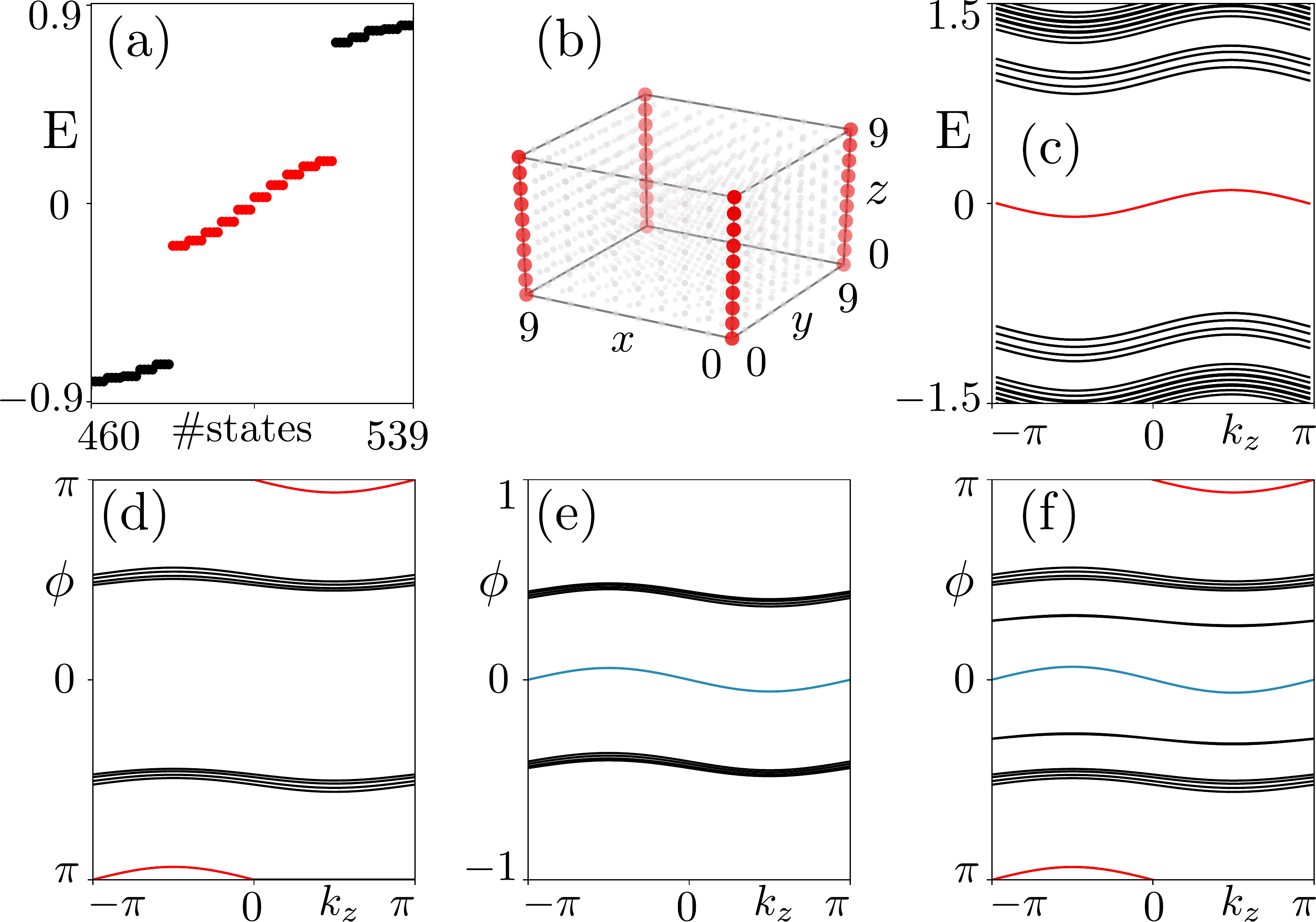}
 \caption{A finite 3D system with $10\times10\times10$ sites in the weak HOTP, with parameters $\gamma_\alpha \equiv0.1$ and $ \lambda_x = \lambda_y = 1$. 
   Panels (a) and (b) show the spectrum and the probability distribution of states forming the  band that crosses $E=0$ (depicted in red). 
  The panels (c)--(f), correspond to a system  which consists of $10\times10$ sites in the \textit{xy} plane and which is infinite along the \textit{z} direction.
 Panel (c) depicts the spectrum of the Hamiltonian as a function of $k_z$.  
 In the remaining panels, we show the eigenphase spectrum of the reflection matrix $r_{\text{2D}}$ for various configurations. 
 In panel (d), we consider an $N_d = 1$ interface. 
 There are two dispersing bands (the mode shown in red is doubly degenerate) that cross $\phi = \pi$ at $k_z = 0, \pi$. 
 In panel (e), the Hamiltonian system with $\gamma_x = 2.1$ is probed with an $N_d = 1$ thick interface. 
 We observe two bands (colored in blue) that cross $\phi = 0$ at $k_z = 0, \pi$. 
 To produce panel (f), we use $N_d = 2$ and a 3D system with gapless hinge modes ($\gamma_x = 0.1$). 
 There are two bands that simultaneously cross $\phi = 0,\pi$ (blue and red).}
\label{fig:HOTI_weak}
\end{figure}

In Fig.~\ref{fig:HOTI_weak}(a), we show the low-energy spectrum of a finite system in the topological phase ($|\gamma_z| = |\gamma_{x,y}| < |\lambda_{x,y}|$) and observe a dispersing mode (shown in red) crossing $E = 0$. 
Looking at the spatial profile of these modes in Fig.~\ref{fig:HOTI_weak}(b), we observe they are located at the hinges.
Extending the 3D system in the $z$ direction while keeping it finite in the \textit{xy} plane, the momentum $k_z$ is a good quantum number, and the dispersion relation along every \textit{z} edge is given by $\gamma_z \sin{k_z}$. 
We plot the spectrum of this system in Fig.~\ref{fig:HOTI_weak}(c), and count four gapless, dispersing hinge bands in total. 
Each band supports two counter-propagating modes per hinge.

We now attach leads to the system that is infinite in the \textit{z} direction as illustrated in Fig.~\ref{fig:sys3d}(b). We assume these leads probe only the surface layer of sites ($N_d = 1$). The eigenphase spectrum of a unitary $r_{\text{2D}}(k_z)$ is plotted in Fig.~\ref{fig:HOTI_weak}(d). 
Because Hamiltonian hinge states are gapless at $k_z = 0, \pi$, the incoming modes of the lead will get reflected with a $\pi$ phase shift at these momenta. 
For this reason, we observe two dispersing bands that cross $\phi = \pi$ at momenta $k_z = 0,\pi$. The states that form these bands are pinned to \textit{z} edges of the lead. 
   
Note that both for the Hamiltonian and for the reflection matrix, the mid gap boundary modes are topologically protected.
Particle-hole symmetry implies that for each Hamiltonian eigenstate and energy $E$ and momentum $k_z$, there must exist an eigenstate at $-E$ and $-k_z$. Similarly, for any eigenstate of the reflection matrix at $\phi$ and $k_z$ there must exist one at $-\phi$ and $-k_z$.
Thus, the boundary-localized mid gap bands cannot be pushed away from $E=0$, or from $\phi=0$ and $\phi=\pi$, without breaking $\mathcal{P}$.

Taking into account all these properties, we define a quantity $\xi_{\rm WTP}^{\pi(0)} (\bar k_z )$ that counts the parity of the number of bands that cross $\phi = \pi (0)$ per \textit{z} edge. Here, $\bar k_z$ denotes (a set of) high-symmetry momenta $\{ 0,\pi \}$. Looking at the spectrum in Fig.~\ref{fig:HOTI_weak}(d), we see that $\xi_{\rm WTP}^{\pi} (\bar k_z) = -1$ and $\xi_{\rm WTP}^{0} (\bar k_z)= 1$.

We now analyze the same scattering setup but with an opposite dimerization ($|\gamma_x| > |\lambda_x|$) in the \textit{x} direction. 
The eigenphase spectrum of $r_{\text{2D}}(k_z)$ is plotted in Fig.~\ref{fig:HOTI_weak}(e) for this case. 
There are two pairs of counter-propagating edge modes crossing $\phi = 0$ at $ k_z = 0,\pi$. As before, each pair is located at a single hinge such that $\xi_{\rm WTP}^{0} (\bar k_z)= -1$ and $\xi_{\rm WTP}^{\pi} (\bar k_z) = 1$. If $|\gamma_{x,y}|> |\lambda_{x,y}|$, the reflection matrix of this 3D system shows no topological features (we opt to not show its $\phi$ spectrum).  

Lastly, we discuss the setup with a thicker interface region ($N_d = 2$) that relates a 3D system with gapless hinge modes with leads. In analogy with Sec.~\ref{sec:corner}, we expect this setup to yield $r_{\text{2D}}$ that simulates a weak anomalous unitary phase. 
The eigenphase spectrum of $r_{\text{2D}}$ is plotted in Fig.~\ref{fig:HOTI_weak}(f). 
We observe gapless states at $\phi = 0,\pi$ for $k_z = 0,\pi$, leading to nontrivial values $\xi_{\rm WTP}^{\pi} (\bar k_z) = -1$ and $\xi_{\rm WTP}^{0} (\bar k_z )= -1$. 

\subsection{3D system with hinge states protected by a mirror symmetry}\label{sec:tcp}

Another way to create hinge modes in 3D systems is to replace dimerization along the \textit{z} direction in 3D BBH model Eq.~\eqref{eq:BBHHam3D} with a non-dimerized nearest-neighbor hopping that preserves the mirror symmetry $\mathcal{M}_z$ at a plane perpendicular to $z$. 
The system then implements a second-order TCP. For concreteness, we consider the momentum space Hamiltonian
\begin{align}\label{eq:BBHHam_TCP}
\begin{split}
h_{\text{TCP}}(\bm{k}) {} &= (\gamma_x  + \lambda_x \cos{k_x}) \eta_z \tau_0 \sigma_x + \lambda_x \sin{k_x} \eta_z \tau_0 \sigma_y \\
& + (\gamma_y  + \lambda_y \cos{k_y}) \eta_x \tau_0 \sigma_0 + \lambda_y \sin{k_y} \eta_y \tau_z \sigma_0 \\
& + (\gamma_z  + \gamma_z \cos{k_z}) \eta_x \tau_x \sigma_z  + 
\gamma_z \sin{k_z} \eta_x \tau_y \sigma_z. 
\end{split}
\end{align}
Besides the local symmetries $\mathcal{P} = \eta_z \tau_0 \sigma_z \mathcal{K}$ and $\mathcal{C} = \eta_z \tau_0 \sigma_z$ (here again, the $\mathcal{C}$ symmetry can be considered as redundant) 
there is also a mirror symmetry $\mathcal{M}_z (k_z) = \eta_0 m_z \sigma_0$, where
 \begin{equation} \label{eq:mirrorTCI}
m_z = 
 \begin{pmatrix}
e^{i k_z} & 0 \\
0 &  1  
\end{pmatrix}.
\end{equation}
This symmetry relates the Hamiltonian at different momenta as
\begin{equation}
\mathcal{M}_z (k_z) h_{\text{TCP}}(k_x, k_y,k_z ) \mathcal{M}^{-1}_z (k_z) = h_{\text{TCP}}(k_x, k_y,-k_z).
\end{equation}
The Hamiltonian is mapped onto itself for two values ($0$ and $\pi$) of momentum $k_z$, such that the mirror symmetry is an element of the little group at these points. This mirror symmetry operator can be written in the form $\mathcal{M}_z (k_z) = V(k_z) D(k_z) V^{\dagger}(k_z)$, where $D(k_z) = \mathop{\rm diag}(1,1,1,1,e^{i k_z}, e^{i k_z}, e^{i k_z}, e^{i k_z})$ is a diagonal matrix of eigenvalues and $V(k_z)$ is a unitary matrix composed of respective eigenvectors. 

At momentum $k_z = \pi$, the mirror symmetry operator has two fourfold degenerate eigenvalues $1$ and $-1$. The rotated Hamiltonian $V(\pi) h_{\text{TCP}}(k_x, k_y,\pi) V^{\dagger}(\pi)$ is block diagonal and consists of two matrix blocks that represent momentum space Hamiltonians of neighboring \textit{xy} layers. These two layers have opposite values of mirror symmetry eigenvalues. For $k_z \neq \pi$, the effective decoupling between two layers is absent because the unitary transformation $V(k_z) h_{\text{TCP}}(k_x, k_y,k_z) V^{\dagger}(k_z)$ does not produce a block-diagonal matrix.

\begin{figure}[tb]
\includegraphics[width=\columnwidth]{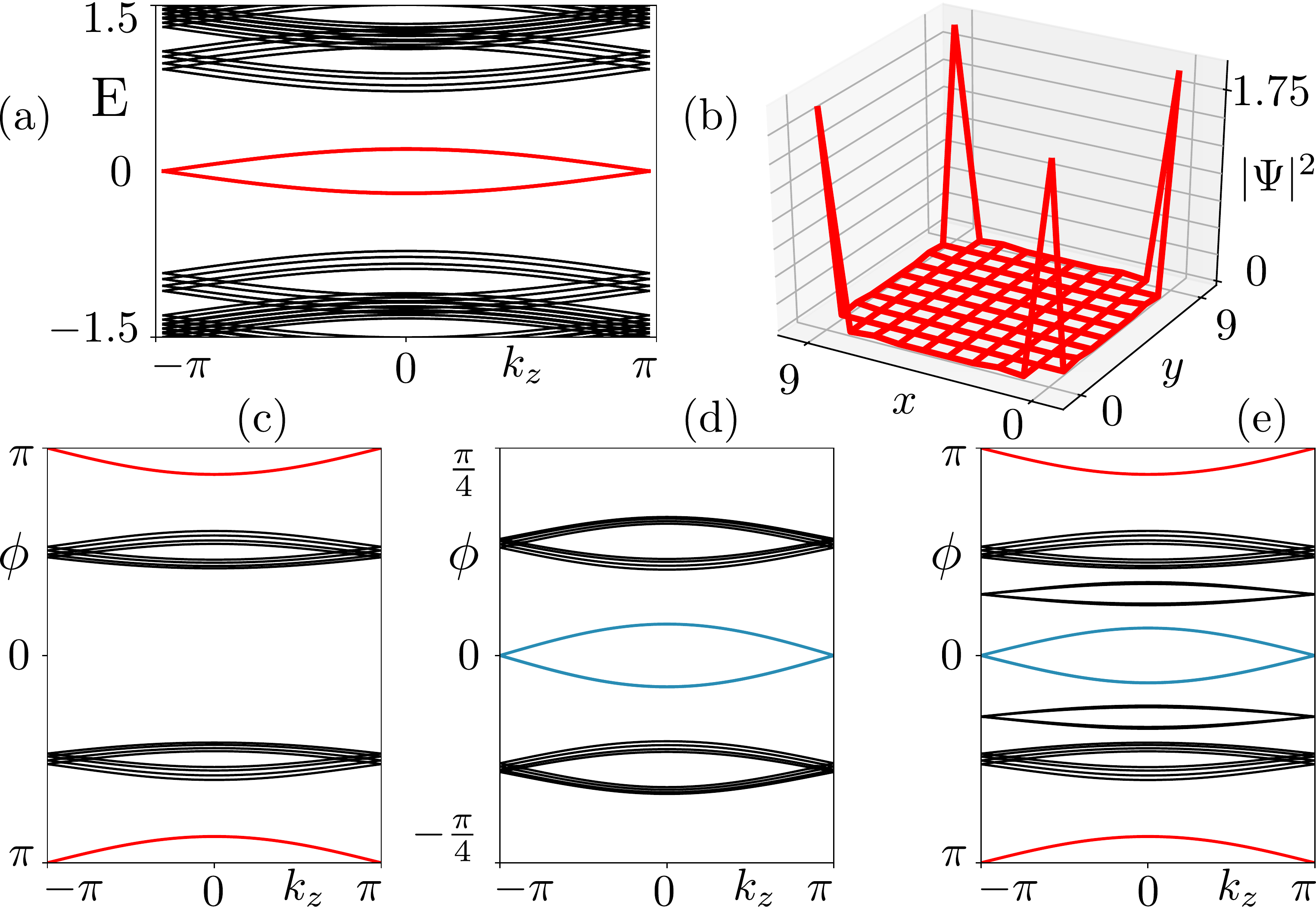}
 \caption{Panel (a) shows the low-energy spectrum of a Hamiltonian system which is infinite in the \textit{z} direction and consists of $10\times10$ sites in the \textit{xy} plane. 
 The bands that cross $E=0$ are shown in red. 
 The sum of the probability distributions of gapless states at $k_z = \pi$ is given in panel (b). 
 In these panels, we use $\gamma_x = \gamma_y = 0.1$, $\lambda_x = \lambda_y = 1$ and $\gamma_z = 0.1$. 
 Panels (c)--(e) show the eigenphase spectra of the reflection matrix $r_{\text{2D}}$. 
 In panel (c), we consider $N_d = 1$ and $\gamma_x = 0.1$. 
 There are four dispersing bands, colored in red, that cross $\phi = \pi$  at $k_z = \pi$. 
 For panel (d), we consider the 3D system with oppositely dimerized \textit{x} edges ($\gamma_x = 2.1$) and $N_d = 1$ and observe four bands (blue) that cross $\phi = 0$  at $k_z=\pi$. In panel (e) the 3D system has gapless hinge modes probed by leads with an $N_d = 2$ thick interface region. We observe the presence of dispersing bands that cross $\phi = 0,\pi$ simultaneously. }
\label{fig:TCI}
\end{figure}

In Fig.~\ref{fig:TCI}(a) is plotted the spectrum of the Hamiltonian for the system that is infinite in the \textit{z} direction and finite in the other two directions. Here, we use $|\gamma_z| = |\gamma_{x,y}| < |\lambda_{x,y}|$. There are eight bands that cross $E = 0$ at $k_z = \pi$ as the coupling between adjacent \textit{xy} layers with amplitude $|\gamma_z| \sqrt{2(1+\cos{k_z})}$ vanishes. In Fig.~\ref{fig:TCI}(b) is plotted the probability distribution of mid gap states at $k_z = \pi$. We conclude that these midgap bands describe states localized at the hinges of a 3D system. 

Next, we calculate the reflection matrix $r_{\text{2D}}$ of the system by attaching two leads as illustrated in Fig.~\ref{fig:sys3d}(b). We start with an $N_d = 1$ thick interface region. The eigenphase spectrum of $r_{\text{2D}}$ is plotted in Fig.~\ref{fig:TCI}(c). 
We observe four dispersing bands crossing $\phi = \pm \pi$ at $k_z =  \pi$. These bands are split into pairs of doubly degenerate bands, and every pair is located at one \textit{z} edge of the 2D system. Therefore, defining a quantity $\xi^{\pi}_{\rm TCP} (\bar k_z = \pi)$ that simply measures the parity of the number of $\pi$ modes per edge is not useful as $\xi^{\pi}_{\rm TCP} (\bar k_z = \pi) = 1$ in this case. It is possible to find an appropriate invariant but not as straightforward as before. Hence, we refrain from doing so here, and direct the reader to Sec.~\ref{sec:char} for more details. 

As before, we proceed by changing the dimerization pattern of the 3D system such that $|\gamma_x| > |\lambda_x|$. The $\phi$ spectrum of $r_{\text{2D}}$ is shown in Fig.~\ref{fig:TCI}(d), where we see four dispersing bands crossing $\phi = 0$ at $k_z = \pi$. 
Finally, we demonstrate that the reflection matrix $r_{\text{2D}}$ can simulate the anomalous Floquet TCP by doubling the thickness of the interface region. 
Its eigenphase spectrum is plotted in Fig.~\ref{fig:TCI}(e). 
We see two sets of bands crossing $\phi = 0, \pi$ at $k_z = \pi$.

\section{Characterization}\label{sec:char}
This Section is devoted to the characterization of the different 2D unitary topological phases described in Sec.~\ref{sec:BBH}. We do this by using topological indices devised for periodically-driven systems, as we have seen in Sec.~\ref{sec:symm} that a unitary reflection matrix obeys the same symmetry conditions as $\mathcal{F}$. Since our approach only provides access to $r_{\rm 2D}$ which corresponds to the time-evolution at the stroboscopic time $T=1$, we will not use topological invariants that require the knowledge of a unitary time-evolution operator $U(t)$ at all times $t$~\cite{Rudner2013, Asboth2014}. 
While such an approach is possible in principle, as shown in Ref.~\cite{Franca2021}, it is more cumbersome than using scattering theory based topological invariants that only need the knowledge of the time evolution at stroboscopic times~\cite{Fulga2016}. 

First, we discuss why scattering theory can be used to identify (static and dynamic) topological phases. Such characterization procedure relies on having a unitary reflection matrix for a scattering region in the topological or trivial phase. This matrix has a different structure in these two phases due to, e.g., a resonant reflection occurring once the incident wave probes a point-like state at the same (quasi-)energy as its own~\cite{Akhmerov2011}. One can then define a quantity that captures these differences and use it as a topological invariant. Importantly, the scattering topological invariant is defined such that it changes its value at the phase transition point of the system it characterizes. At this point, the reflection matrix is singular as a result of a nonzero conductivity between the leads enabled by a gap closing in the scattering region. 
 
The above-mentioned requirement of a unitary reflection matrix for a characterization procedure is automatically obeyed in the trivial phase that does not conduct in the bulk nor at its boundaries. For a system in the topological phase, whether the reflection matrix is unitary or not depends on the order of the topological phase and what boundary conditions are applied. In the case of STPs, we direct the interested reader to Ref.~\cite{Fulga2012}. Here, we focus on 2D systems realized in Sec.~\ref{sec:BBH} that have topological states either at the corners or at a pair of \textit{z} edges. Since these states are not present at all boundaries of the system, a suitable transport experiment that detects a unitary $r$ may be designed with a finite-sized system. For the reasons explained later, we call the leads that probe a unitary topological phase absorbing terminals. 

\begin{figure}[tb]
\includegraphics[width=1\columnwidth]{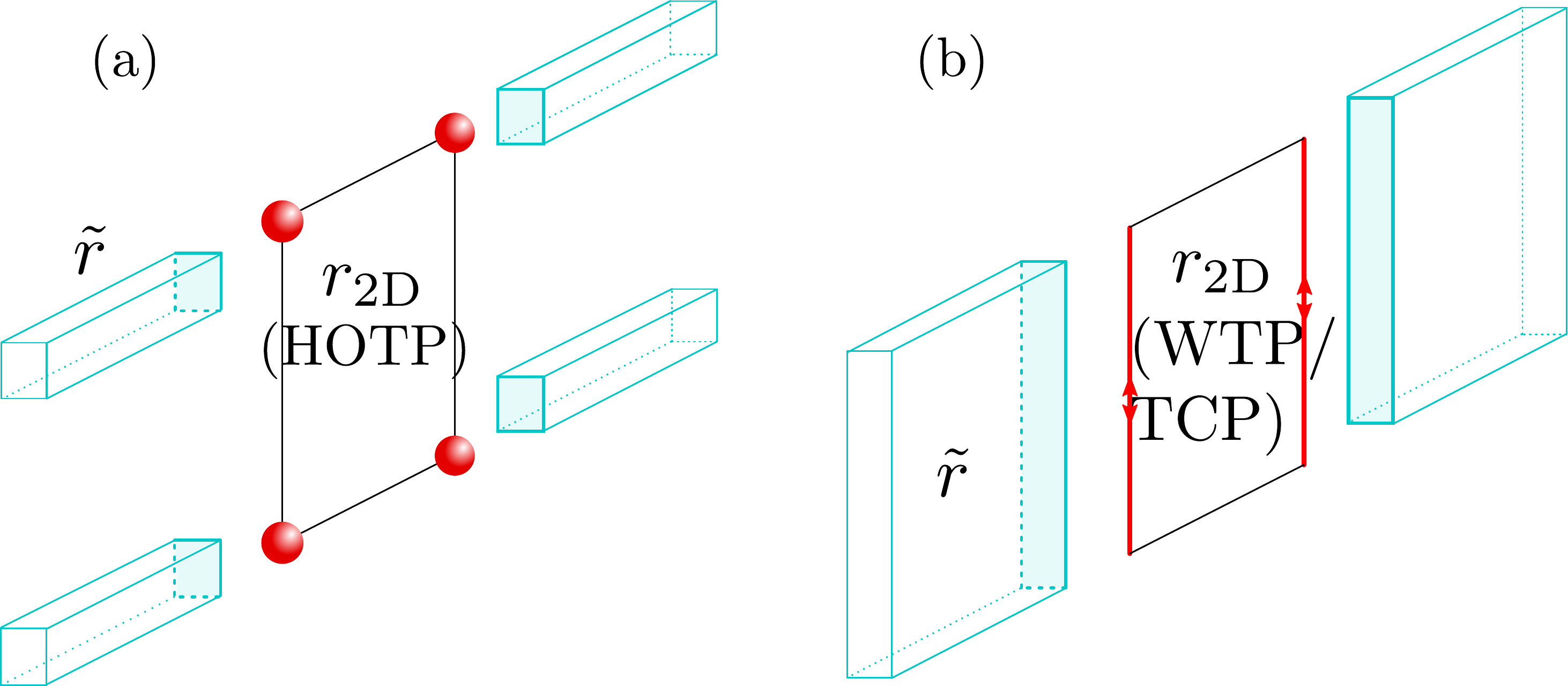}
\caption{Sketch of the absorbing terminals (blue) used to characterize 2D unitary systems.
 In panel (a) are shown four point-like terminals that probe the corners of a 2D system described by $r_{\rm 2D}$. In panel (b), the terminals are placed on \textit{z} edges that also host propagating topological modes of a 2D system described by $r_{\rm 2D}$. }
\label{fig:3Dscat_setup}
\end{figure}

For a unitary second-order topological phase, appropriate absorbing terminals are illustrated in Fig.~\ref{fig:3Dscat_setup}(a). They are point-like and located at all four corners of the system such that they directly probe the corner states. In this way, it is possible to detect the bulk and edge gap closing(s) of a 2D system that separate a topological from a trivial phase~\cite{Benalcazar2017, Bomentara2019, Rodriguez-Vega2019}. For unitary systems in Sec.~\ref{sec:hinge} and~\ref{sec:tcp}, suitable terminals are placed on those edges where dispersive, topological edge states are present, as illustrated in Fig.~\ref{fig:3Dscat_setup}(b). Note that every terminal is designed such that it probes four corner sites in the case of Fig.~\ref{fig:3Dscat_setup}(a), and two outermost layer of sites in the case of  Fig.~\ref{fig:3Dscat_setup}(b). 

In the following, we show how the scattering matrix of a unitary system is calculated, and then we discuss topological invariants for all three cases considered in Sec.~\ref{sec:BBH}. 

\subsection{Nested scattering matrix procedure}\label{sec:nested}

To calculate the scattering matrix of the unitary system, we follow the procedure outlined in Ref.~\cite{Fulga2016}. It is based on defining fictitious terminals assumed to absorb particles at the stroboscopic times. These terminals are positioned at the boundaries of interest, like the ones shown in Fig.~\ref{fig:3Dscat_setup}. The boundary consists of $N_T$ sites with $N_d $ degrees of freedom per site (following the convention outlined in Sec.~\ref{subsec:Scatt_setup}). Mathematically, the terminals are described with the rectangular matrix $P$ that has $N_T \times N$ entries of size $N_d \times N_d$, where $N$ denotes the total number of sites of the unitary system. This matrix maps each site $i \in \{1,\dots,N_T\}$ on the boundary to its site $j\in\{1,\dots, N\}$ in the Floquet system, \emph{i.e.}, $P_{ij} = \mathbbm{1}_{N_d}$ if $i$ and $j$ are the same physical site and $P_{ij}=0_{N_d}$ otherwise. 

In analogy to the possibility of calculating the scattering matrix of a static system at any energy $E$ (in this work, we have used only $E=0$), the scattering matrix of a unitary system with eigenphases $\phi_n$ can be determined at any phase $\phi \in[0,2\pi )$ in the following way
\begin{equation}\label{eq:new_scat_matrix}
  \tilde{S}(\phi) = P [\mathbbm{1}_{N\times N_d} - e^{-i \phi} r_{\rm 2D} (\mathbbm{1}_{N\times N_d} - P^T P)]^{-1}  e^{-i \phi} r_{\rm 2D} P^T.
\end{equation}
For details on how this formula is obtained and its connection with periodically driven systems, see Appendix \ref{app:Floquet_smatrix}.

Since the scattering matrix $\tilde{S}(\phi)$ is calculated from the reflection matrix $r_{\rm 2D}$ alone, we dub this procedure the method of nested scattering matrices~\cite{Franca2021}. 
From now on, all nested scattering matrices and their sub-blocks are denoted with a tilde. The phase $\phi \in[0,2\pi)$ corresponds to an additional phase $e^{-i\phi}$ accumulated during the time $T$. 
In this sense, it serves as a simple generalization of the stroboscopic time-evolution which at $\phi=0$ is simply given by $r$. In the following, we will be mainly interested at the values $\phi=0$ (no additional phase) and $\phi=\pi$ (additional minus sign). For simplicity, we denote these specific phases ($0$ and $\pi$) as $\bar \phi$.

\subsection{Topological invariants}\label{sec:top_inv}

Scattering topological invariants are defined using local symmetry constraints imposed on the reflection matrix $\tilde{r}(\phi)$ that is a block of $\tilde{S}(\phi)$. These constraints can be obtained by combining Eq.~\eqref{eq:S_symm} and Eq.~\eqref{eq:new_scat_matrix}. For example, the reflection matrix $\tilde{r}(\phi)$ of particle-hole symmetric unitary systems obeys
\begin{equation} \label{eq:rtilde_PH}
U_{\mathcal{P}} \tilde{r}( \phi)^* U_{\mathcal{P}}^{\dagger} = \tilde{r}(-\phi).
\end{equation}
Note that the size of a block diagonal matrix $U_{\mathcal{P}}$ used above differs from the one in Eq.~\eqref{eq:S_symm}. Here, it has $\frac{N_T}{2} \times \frac{N_T}{2}$ entries of size $N_d \times N_d$. 

The Eq.~\eqref{eq:rtilde_PH} implies that for $\phi=\bar\phi \in \{0,\pi\}$ the matrix $\tilde{r}$ has a real determinant. Moreover, the reflection matrix $\tilde{r}( \bar \phi)$ of all unitary topological phases studied in Sec.~\ref{sec:BBH} is unitary, provided they are probed with appropriate fictitious scattering setups. For this reasons, the determinant of $\tilde{r}( \bar \phi)$ can only take values $1$ and $-1$. We shall see later that this property was used to define the scattering topological invariant of a particle-hole symmetric system. 

For topological phases protected by chiral symmetry, the constraint on the reflection matrix reads
\begin{equation}\label{eq:top_inv_chiral}
U_{\mathcal{C}} \tilde{r}(\phi)^{\dagger} U_{\mathcal{C}}^{\dagger} = \tilde{r}(-\phi).
\end{equation}
As before, the sizes of a block diagonal matrix $U_{\mathcal{C}}$ are different in Eqs.~\eqref{eq:top_inv_chiral} and~\eqref{eq:S_symm}.

The relation Eq.~\eqref{eq:top_inv_chiral} implies that for $\phi = \bar\phi$, it is possible to find a basis in which $\tilde{r}(\phi)$ is Hermitian. These specific phases also render matrix $\tilde{r}(\phi = \bar\phi)$ unitary for the topological phases studied in this work. Hence, the eigenvalues of this matrix can only take values $1$ and/or $-1$, and the topological invariant $\nu^{\bar\phi}_C$ can be defined as the number of negative eigenvalues of $\tilde{r}(\bar\phi)$~\cite{Fulga2011, Fulga2012}.

Lastly, we emphasize once more that for the models we have studied, it is not necessary to calculate both particle-hole and chiral topological invariants. The value of the chiral invariant can be inherited from the value of the particle-hole invariant (or vice versa) because our static scattering regions support a minimal number of modes per $(D-n)$-dimensional boundaries. The chiral invariant equals $1$ whenever the $\mathcal{P}$ invariant takes a nontrivial value. For this reason, we do not write explicitly its value in the following.

\subsubsection{Unitary second-order topological phase}\label{subsec:corner_inv}

The unitary second-order topological phases studied in Sec.~\ref{sec:corner} can be characterized with corner terminals illustrated in Fig.~\ref{fig:3Dscat_setup}(a). 
It is possible to define two $\Z_2$ topological invariants as~\cite{Fulga2016}
\begin{equation}\label{eq:top_inv}
\nu^0 = \mathop{\rm sgn} \det[\tilde{r} (\bar \phi = 0)], \; 
\nu^{\pi} = \mathop{\rm sgn} \det[\tilde{r}(\bar \phi = \pi)].
\end{equation}

We choose the convention that $\nu^{\bar\phi} =+1$ ($\nu^{\bar\phi} = -1$) corresponds to the trivial (nontrivial) phase. The origin of the difference in values of $\nu^{\bar\phi}$ is the aforementioned resonant reflection occuring once the phase at which terminals probe the system has the same value as the eigenphase of a topologically protected mode. The invariant $\nu^{0,\pi}$ is therefore in correspondence to the quantity $\xi^{0,\pi}$ introduced in Sec.~\ref{sec:corner}, that counted the parity of the number of modes per corner at eigenphases $\phi =  0, \pi$. We verified that whenever $\xi^{0,\pi} = -1 $ so is $\nu^{0,\pi} = -1$.

\subsubsection{Unitary weak topological phase}\label{subsec:weak_inv}

As illustrated in Fig.~\ref{fig:3Dscat_setup}(b), a WTP is characterized with 1D absorbing terminals. The scattering topological invariant is calculated at the high-symmetry points $\bar k_z \in \{0,\pi\}$ and reads~\cite{Ladovrechis2019}
\begin{equation}\label{eq:weak_inv}
\nu_{\text{WTP}}^{\bar\phi} (\bar k_z) = \mathop{\rm sgn} \det[\tilde{r} (\bar \phi, \bar k_z)].
\end{equation}

Like previously, we may relate this invariant to a quantity $\xi_{\rm WTP}^{\pi(0)} (\bar k_z )$ that measures the parity of the number of hinge states, see Sec.~\ref{sec:hinge}. Whenever $\xi_{\rm WTP}^{\pi(0)} (\bar k_z )= -1$, the invariant $\nu_{\text{WTP}}^{\bar\phi} (\bar k_z) = -1$ due to a resonant reflection.  
This prediction was tested and verified for all the matrices $r_{\rm 2D}$ whose spectra are shown in Fig.~\ref{fig:HOTI_weak}(c-e).  

\subsubsection{Unitary mirror-protected topological phase}\label{subsec:tcp_inv}

To characterize this unitary topological phase, we place the absorbing terminals as shown in Fig.~\ref{fig:3Dscat_setup}(b). Since every \textit{z} edge hosts a pair of propagating topological modes that become gapless at $k_z = \pi$, the invariant Eq.~\eqref{eq:weak_inv} calculated at this momentum gives a trivial result. This is in agreement with having $\xi_{\text{TCP}}^{\bar\phi} (\bar k_z) = 1$. 

To find the invariant capable of capturing this kind of topology, we remember that neighboring \textit{xy} layers of the Hamiltonian scattering region are effectively decoupled at momentum $k_z = \pi$ due to mirror symmetry $\mathcal{M}_z$. Therefore, an incoming plane wave incident to one layer remains confined to this layer until it gets absorbed back into the lead. Consequently, the unitary system given by $r_{2D}(y, k_z = \pi)$ describes two decoupled chains (with $N_d$ degrees of freedom per site) spanned in the \textit{y} direction. Following the same logic for the fictitious scattering problem, the reflection matrix $\tilde{r}$ of this unitary system can be written in the basis in which it is a block diagonal matrix
\begin{equation} \label{eq:refl_blocks_TCI}
 \tilde{r} (\bar \phi,\bar k_z = \pi) = 
 \begin{pmatrix}
\tilde{r}^{+} & 0 \\ 0 & \tilde{r}^{-}  
\end{pmatrix}.
\end{equation}
Here, $\tilde{r}^{\pm}$ denote reflection matrices of two layers with opposite mirror symmetry eigenvalues, as discussed in Sec.~\ref{sec:tcp}. 

At $\bar k_z = \pi$, the particle-hole symmetry leads to the 
constraint~\cite{Ladovrechis2019}
\begin{equation}\label{eq:r_restr_TCI}
U_{\mathcal{P}} \tilde{r} (\phi, \pi) U_{\mathcal{P}} = \tilde{r}^{*} (-\phi, \pi),
\end{equation}
which allows us to define a $\Z_2$ invariant
\begin{equation}\label{eq:TCI_inv}
\nu_{\text{TCP}}^{\bar\phi, \pm} = \mathop{\rm sgn} \det[\tilde{r}^{\pm} ( \bar\phi, \pi)].
\end{equation}

We now calculate this quantity numerically, and find that for a system shown in Fig.~\ref{fig:HOTI_weak}(c), $\nu_{\text{TCP}}^{0, \pm} = 1$ and $\nu_{\text{TCP}}^{\pi, \pm} = -1$ thus capturing the presence of dispersive modes at $\phi = \pi$. For the system with the eigenphase spectrum given in Fig.~\ref{fig:HOTI_weak}(d), the invariants read $\nu_{\text{TCP}}^{0, \pm} = -1$ and $\nu_{\text{TCP}}^{\pi, \pm} = 1$. Finally, the existence of the anomalous unitary phase in Fig.~\ref{fig:HOTI_weak}(f) is confirmed by nontrivial values of $\nu_{\text{TCP}}^{0, \pm} = -1$ and $\nu_{\text{TCP}}^{\pi, \pm} = -1$.

\section{Conclusion} \label{sec:conclusion}

Equilibrium systems with gapless corner states present an opportunity to realize unitary Floquet topological phases in a way that circumvents some of the problems connected to their experimental realization. 
Namely, even though Floquet phases result from periodic driving, we have found they can be simulated by unitary reflection processes of the systems supporting HOTPs. 
In this way, the absence of a driving field eliminates noise-induced decoherence.

Here, we have provided examples of unitary topological phases protected by the combination of local and crystalline symmetries, thus expanding the range of Floquet phases initially realized with this approach~\cite{Franca2021}. 
First, we have studied 3D systems supporting a HOTP, showing that their reflection matrices describe 2D systems with corner modes at eigenphases $0$ and/or $\pi$. Other kinds of interesting 3D systems to study are the ones with hinge modes protected either by translation symmetry or mirror symmetry. The reflection matrix from the surface bounded by these hinge modes is unitary. It supports edge states at eigenphases $0$ and/or $\pi$ thus simulating a first-order Floquet WTP and TCP, respectively. For all phases, we have defined and calculated the topological invariants based on scattering theory.

While we have focused exclusively on uniform, disorder free systems in this work, we expect our results to remain robust against the addition of impurities.
For unitary phases with corner states, we have shown previously that the topological phase remains intact upon adding particle-hole symmetric disorder, provided that the bulk gap does not close \cite{Franca2021}.
Further, weak and crystalline topological phases have been shown to be robust against disorder provided that the protecting lattice symmetries are still preserved on average, both in Hermitian \cite{Fulga2014, Diez2014} as well as unitary systems \cite{Fulga2016, Fulga2019} .

There are several research directions open for future works. 
For example, one can study reflection matrices of systems in HOTPs where gapless corner states are not present at every corner~\cite{Franca2019, Zhu2018, Volpez2019}. 
This could possibly simulate unitary phases with an odd number of topologically protected modes. 
It would be also interesting to explore how these setups can be realized 
either in superconductors, where the same formalism applies provided that superconductivity is treated at the mean field level \cite{Qi2011}, or
in meta-materials such as topoelectric circuits~\cite{Imhof2018} or photonic systems~\cite{Noh2018}. 
Another possibility would be to explore how our mapping relates classification tables of static and Floquet phases.

\section{Acknowledgments} \label{sec:acknowledgement}
We thank Ulrike Nitzsche for technical assistance. This work was supported by the Deutsche Forschungsgemeinschaft (DFG, German Research Foundation) through the W\"{u}rzburg-Dresden Cluster of Excellence on Complexity and Topology in Quantum Matter -- \emph{ct.qmat} (EXC 2147, project-id 390858490) and under Germany's Excellence Strategy -- Cluster of Excellence Matter and Light for Quantum Computing (ML4Q) EXC 2004/1 390534769.

\appendix
\section{Analytical reflection matrix} \label{app:R_Green}

In this appendix, we analytically calculate the reflection matrix $r_{\text{2D}}$ using the boundary Green's function~\cite{Peng2017} and the Mahaux-Weidenm{\"u}ller formula~\cite{Mahaux1969}.
We start from a real space version of the 3D tight-binding BBH model given in Eq.~\eqref{eq:BBHHam3D}
\begin{align} \label{eq:RealBBH}
  &H \!=\!  - \!\!\!\sum_{m,n,l}\! (-1)^{n+l} (\gamma_x c_{2m,n,l}^{\dagger} c_{2m-1,n,l}^{}\! +\! \lambda_x c_{2m + 1,n,l}^{\dagger} c_{2m,n,l}^{} )\nonumber \\  
& + \sum_{m,n,l} (\gamma_y c_{m,2n,l}^{\dagger} c_{m,2n-1,l}^{} + \lambda_y c_{m,2n+1,l}^{\dagger} c_{m,2n,l}^{} )  \nonumber\\  
& + \sum_{m,n,l} (-1)^{n} (\gamma_z c_{m,n,2l}^{\dagger} c_{m,n,2l-1}^{} + \lambda_z c_{m,n,2l+1}^{\dagger} c_{m,n,2l}^{} ) \nonumber\\
&+ \text{H.c}. 
\end{align}
Some hoppings in the \textit{x} and \textit{z} direction are negative, in order to realize a magnetic $\pi$ flux on each face of the unit cell. 
As in Eq.~\eqref{eq:BBHHam3D}, we assume that all hopping amplitudes are real. 

In the following, we derive a fixed-point boundary Green's function for the \textit{yz} surface using the transfer matrix technique~\cite{Peng2017}.  
To this end, we introduce a spinor
\begin{equation*}
  \Psi_{m,n,l} = (c_{m,2n-1,2l}, c_{m,2n,2l-1}, c_{m,2n,2l}, c_{m,2n-1,2l-1}), 
\end{equation*}
where the indices $m,n,l$ assume the values $m = 1,\dots,L$, $n = 1,\dots,W/2$ and $l = 1,\dots,H/2$. 
For periodic boundary conditions in the \textit{y} and \textit{z} directions, its Fourier transform in the $(k_y, k_z)$ space reads
\begin{equation*}
\Psi_{m} (k_y, k_z) = \frac{2}{\sqrt{W H}} \sum_{k_y} \sum_{k_z} 
e^{i n k_y + i l k_z} \Psi_{m,n,l}. 
\end{equation*}

Then, the Hamiltonian Eq.~\eqref{eq:RealBBH} can be written as
\begin{align*}
\begin{split}
  H = \sum_{k_y, k_z} \Biggl[ & \sum_{m = 1}^{L} \Psi_{m}^{\dagger} (k_y, k_z) h_m(k_y,k_z) \Psi_{m} (k_y, k_z) +  \\
  & \sum_{m = 1}^{L-1}  \Psi_{m+1}^{\dagger} (k_y, k_z) V_m \Psi_{m} (k_y, k_z) + \text{H.c} \Biggr],
\end{split}
\end{align*} 
where 
\begin{align*}\label{eq:2DsliceBBH}
\begin{split}
h_m (k_y, k_z) = & (\gamma_y + \lambda_y \cos{k_y})\mu_x \nu_0 + \lambda_y \sin{k_y} \mu_y \nu_z \\
& +  (\gamma_z + \lambda_z \cos{k_z}) \mu_y \nu_y + \lambda_z \mu_y \nu_x,
\end{split}
\end{align*}
and 
\begin{equation*}\label{eq:Xhopp}
  V_{m} =
    \begin{cases}
      \gamma_x \mu_z \nu_0, & \text{$m$ is odd},\\
      \lambda_x \mu_z \nu_0, & \text{ $m$ is even}.
    \end{cases}       
\end{equation*}
Here, matrices $\mu$ and $\nu$ act on the sublattice sites in the 2D plane. 

The transfer matrix $M_m$ (in the $x$ direction) is given by
\begin{eqnarray}\label{eq:tmatrix}
M_m (\omega) =\begin{pmatrix} (V_m^{\dagger})^{-1} g_m(\omega)^{-1} & - (V_m^{\dagger})^{-1} V_{m-1} \\ \mathbbm{1}_4 & 0 \end{pmatrix},
\end{eqnarray}
where $g^{-1}_m (\omega) = \omega \mathbbm{1}_4 - h_m (k_y,k_z)$ and $\mathbbm{1}_4$ is a $4\times 4$ unit matrix. 
Following the procedure outlined in Ref.~\cite{Peng2017}, we only consider $\omega = 0$ in the following.
Due to the staggering, the transfer matrix $T$ across a unit cell is given by the product $T= M_2 M_1$. We obtain
\begin{widetext}
\begin{align}
\begin{split}
T = 
\left(
\begin{array}{cccccccc}
 \chi & 0 & 0 & 0 & 0 & 0 & \frac{\gamma _y+e^{-i k_y} \lambda _y}{\gamma _x} & -\frac{\gamma _z+e^{i k_z} \lambda _z}{\gamma _x} \\
 0 & \chi & 0 & 0 & 0 & 0 & \frac{\gamma _z+e^{-i k_z} \lambda _z}{\gamma _x} & \frac{\gamma _y+e^{i k_y} \lambda _y}{\gamma _x} \\
 0 & 0 & \chi & 0 & -\frac{\gamma _y+e^{i k_y} \lambda _y}{\gamma _x} & -\frac{\gamma _z+e^{i k_z} \lambda _z}{\gamma _x} & 0 & 0 \\
 0 & 0 & 0 & \chi & \frac{\gamma _z+e^{-i k_z} \lambda _z}{\gamma _x} & -\frac{\gamma _y+e^{-i k_y} \lambda _y}{\gamma _x} & 0 & 0 \\
 0 & 0 & -\frac{\gamma _y+e^{-i k_y} \lambda _y}{\gamma _x} & \frac{\gamma _z+e^{i k_z} \lambda _z}{\gamma _x} & -\frac{\lambda _x}{\gamma _x} & 0 & 0 & 0 \\
 0 & 0 & -\frac{\gamma _z+e^{-i k_z} \lambda _z}{\gamma _x} & -\frac{\gamma _y+e^{i k_y} \lambda _y}{\gamma _x} & 0 & -\frac{\lambda _x}{\gamma _x} & 0 & 0 \\
 \frac{\gamma _y+e^{i k_y} \lambda _y}{\gamma _x} & \frac{\gamma _z+e^{i k_z} \lambda _z}{\gamma _x} & 0 & 0 & 0 & 0 & -\frac{\lambda _x}{\gamma _x} & 0 \\
 -\frac{\gamma _z+e^{-i k_z} \lambda _z}{\gamma _x} & \frac{\gamma _y+e^{-i k_y} \lambda _y}{\gamma _x} & 0 & 0 & 0 & 0 & 0 & -\frac{\lambda _x}{\gamma _x} \\
\end{array}
\right),
\end{split}
\end{align}
 \end{widetext}
where $\chi = -\frac{\gamma _x^2+\gamma _y^2+\gamma _z^2+\lambda _y^2+\lambda _z^2 + 2 \gamma _y \lambda _y \cos \left(k_y\right)+2 \gamma _z \lambda _z \cos \left(k_z\right)}{\gamma _x \lambda _x}$. 

\begin{figure*}[tb]
\includegraphics[width=1\textwidth]{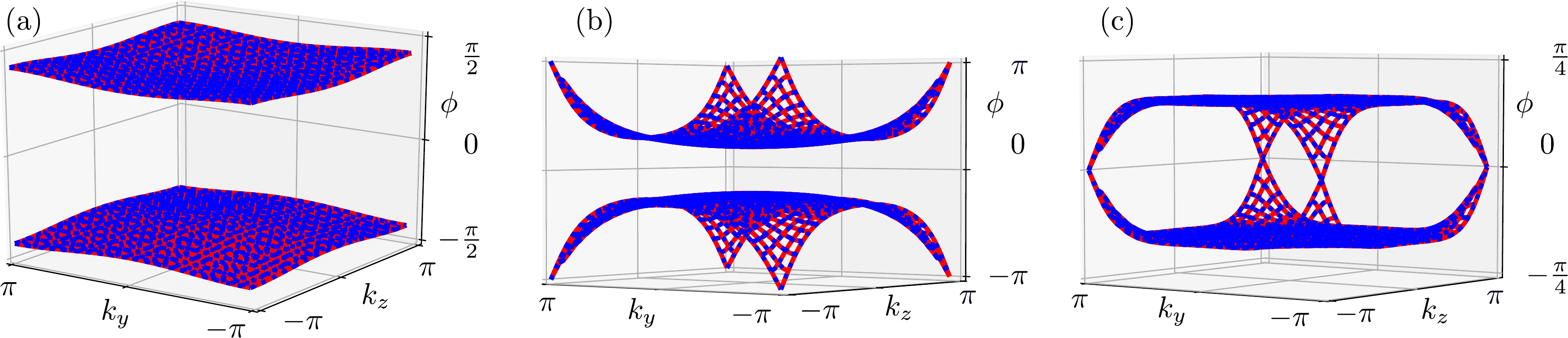}
 \caption{Eigenphase spectrum of $r_{\text{2D}}$ for a system that is infinite in both the \textit{y} and \textit{z} directions. 
 The bands obtained using the analytical approach are shown in blue, and the numerical results are shown in red. 
 In panel (a), the parameters are $\gamma_\alpha = 0.1$ and $\lambda_\alpha =1$. 
 In panels (b) and (c), we show results for $\gamma_{y,z}  = 0.999$ and $\lambda_{\alpha}=1$. 
 For panel (b), $\gamma_x = 0.1$, while $\gamma_x = 2.1$ in panel (c). }
\label{fig:comp_AN}
\end{figure*}

We are interested in the eigenvalues $\Lambda_m$ of $T$ which describe the attenuation along $x$. For $\gamma_y = \gamma_ z= \gamma$ and $\lambda_y = \lambda_z = \lambda$ they are given by
\begin{equation}
\Lambda_{1,2} = \frac{\chi}{2} - \frac{\lambda_x}{2\gamma_x} \mp \sqrt{\Bigl(\frac{\chi}{2} - \frac{\lambda_x}{2\gamma_x}\Bigr)^2 - 1},
\end{equation}
each four-times degenerate.
Due to the fact that $T$ is symplectic (which corresponds to the current conservation), the
eigenvalues obey the relation $\Lambda_ 1 = 1/\Lambda_2^*$. We collect all the eigenvalues with $|\lambda|>1$ into the diagonal matrix $\Lambda$ and decompose $T$ into
\begin{equation}
T =  \begin{pmatrix}
U_{11} & U_{12} \\ U_{21} & U_{22} 
\end{pmatrix} 
\begin{pmatrix} \Lambda & 0 \\ 0 & 1/\Lambda^{*} \end{pmatrix}
\begin{pmatrix}
U_{11} & U_{12} \\ U_{21} & U_{22} 
\end{pmatrix}^{-1}. 
\end{equation}

The fixed-point boundary Green's function is found using the relation 
\begin{equation}
G = U_{21} U_{11}^{-1} (V_0^{\dagger})^{-1},
\end{equation}
where $V_0 = \lambda_x \mu_z \nu_0$. It reads

\begin{widetext}
\begin{align}
\begin{split}
G(k_y,k_z) = \left(
\begin{array}{cccc}
 0 & 0 & \frac{\gamma +\lambda  e^{-i k_y}}{\lambda _x \left(\gamma _x \lambda_1 +\lambda _x\right)} & -\frac{\gamma +\lambda  e^{i k_z}}{\lambda _x \left(\gamma _x \lambda_1 +\lambda _x\right)} \\ 
 0 & 0 & \frac{ \gamma +\lambda  e^{-i k_z}}{\lambda _x \left(\gamma _x \lambda_1 +\lambda _x\right)} & \frac{ \gamma +\lambda  e^{i k_y}}{\lambda _x \left(\gamma _x \lambda_1 +\lambda _x\right)} \\ 
 \frac{ \gamma +\lambda  e^{i k_y}}{\lambda _x \left(\gamma _x \lambda_1 +\lambda _x\right)} & \frac{ \gamma +\lambda  e^{i k_z}}{\lambda _x \left(\gamma _x \lambda_1 +\lambda _x\right)} & 0 & 0 \\
 -\frac{ \gamma + \lambda  e^{-i k_z}}{\lambda _x \left(\gamma _x \lambda_1 +\lambda _x\right)} & \frac{ \gamma +\lambda  e^{-i k_y}}{\lambda _x \left(\gamma _x \lambda_1 +\lambda _x\right)} & 0 & 0 
\end{array}
\right).
\end{split}
\end{align}
 \end{widetext}
 
Using the Mahaux-Weidenm{\"u}ller formula, the reflection matrix $r_{\text{2D}}^{\text{an}} (k_y, k_z)$ can be expressed as~\cite{Peng2017} 
\begin{equation}\label{eq:MW}
r_{\text{2D}}^{\text{an}} (k_y, k_z) = \frac{\mathbbm{1}_4 - i W G(k_y,k_z) W^{\dagger}}{\mathbbm{1}_4 + i W G(k_y,k_z) W^{\dagger}}.
\end{equation}
Here, matrix $W$ represents coupling of the lead to the system. 
In this work, we assume it is a diagonal matrix with unit entries, \emph{i.e.}, $W = \mathbbm{1}_4$. 
In this case, Eq.~\eqref{eq:MW} can be simplified to
\begin{equation}\label{eq:MW_simp}
r_{\text{2D}}^{\text{an}} (k_y, k_z) = \frac{\mathbbm{1}_4 - i G(k_y,k_z)}{\mathbbm{1}_4 + i G(k_y,k_z)}
\end{equation}
such that $r_{\text{2D}}^{\text{an}}$ is the Caley transform of $G$.
Note that a weak link between the leads and the system would correspond to $W = \mathbbm{1}_4/t_{ls}$. 

In the following, we compare the eigenphase spectrum of $r_{\text{2D}}^{\text{an}}$ with the numerically obtained results~\cite{codeKwant}. 
The results are plotted in Fig.~\ref{fig:comp_AN} for different values of $\gamma$ and $\lambda$. Both approaches yield the same eigenphase spectra.
In Fig.~\ref{fig:comp_AN}(a), we show the $\phi$ spectrum for a 3D system in the HOTP. 
We see that the bulk of the 2D system, described by $r_{\text{2D}}^{\text{an}}$ is gapped. 
In Fig.~\ref{fig:comp_AN}(b) and Fig.~\ref{fig:comp_AN}(c), we study topological phase transitions in the bulk of the 2D system described by a unitary $r_{\text{2D}}^{\text{an}}$. For $|\gamma_x|<|\lambda_x|$, the bands close the gap at $\phi = \pi$ for $|\gamma_{y,z}| = |\lambda_{y,z}|$. If the 2D system is finite in the \textit{y} and \textit{z} directions, this phase transition would lead to a hybridization of $\pi$ modes with the bulk~\cite{Franca2021}. Once $|\gamma_x|>|\lambda_x|$ and $|\gamma_{y,z}| < |\lambda_{y,z}|$, the finite 2D system supports $0$ modes at its corners, see Fig.~\ref{fig:3DHOTI_r}(b). The existence of these modes is related to the nontrivial $r_{\text{2D}}(k_y, k_z)$ bulk gap around $\phi = 0$ that gets closed for $|\gamma_{y,z}| = |\lambda_{y,z}|$ as shown in Fig.~\ref{fig:comp_AN}(c).

\section{Transversal lead coupling} \label{app:lead_coupling}

So far, we have calculated the reflection matrices using a model of idealized leads consisting of decoupled waveguides. 
This approximation was used because we expect a high degree of control over all degrees of freedom in meta-material platforms that were mostly used to experimentally realize HOTPs. 
It would be, however, prudent to verify that our conclusions remain intact in the presence of a transversal coupling between neighboring waveguides in the plane parallel to system-lead interface.

To simulate this effect, we allow for a nonzero coupling $t_p$ between neighboring waveguides, which is present only on the first 10 sites of the lead, and which has equal amplitude in both the $y$ and $z$ directions. 
Thus, $t_p$ will mix the incoming and outgoing modes of different chains and alter the reflection matrix.

\begin{figure}[tb]
\includegraphics[width=1\columnwidth]{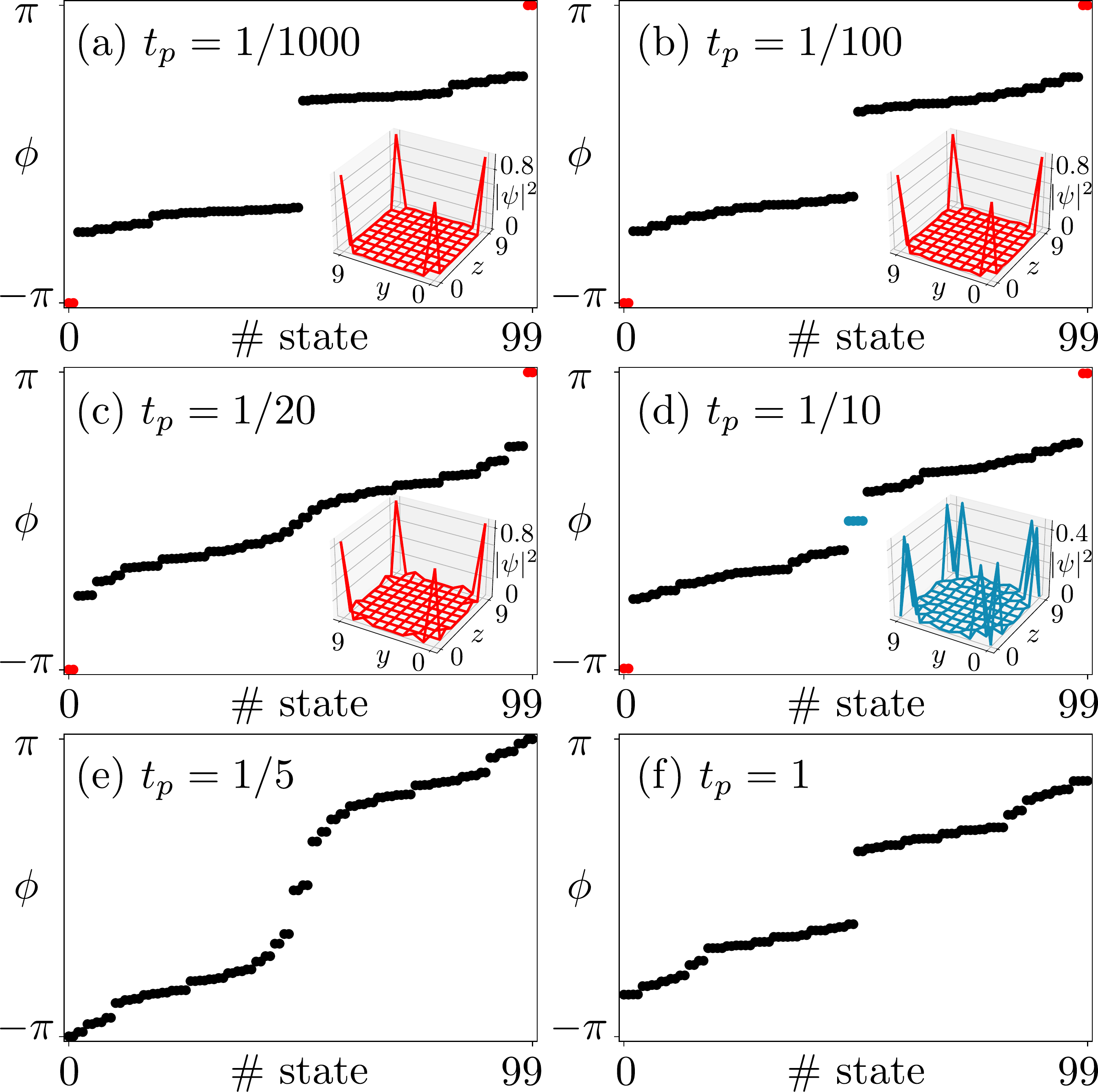}
 \caption{Eigenphase spectrum of $r_{2D}$ and probability density of corner states as a function of the transverse coupling $t_p$. 
 The scattering region contains $L^3$ sites, where $L = 10$, and $t_p$ couples waveguides within the first 10 sites from the scattering region. 
 Other parameters read: $\gamma_{\alpha} = 0.1$, $\lambda_{\alpha} = 1$, $t_l = 1$. 
 The probability densities of midgap modes are plotted only if their eigenphases differ from $\phi=0,\pi$ by less than $0.01$.  }
\label{fig:non_ideal_leads}
\end{figure}

In the following, we show results for different strengths of the hopping $t_p$ on the reflection matrix of the 3D BBH model with zero-energy corner states. 
The spectrum of this scattering region and probability density of zero-energy states are shown in Fig.~\ref{fig:3DHOTI} for $t_p=0$. 
The effect of nonzero $t_p$ is shown in Fig.~\ref{fig:non_ideal_leads}.

When $t_p$ is negligible compared to $t_l = 1$ (the hopping along the 1D waveguides), we expect that the inter-waveguide coupling has little effect on the topological properties of the reflection matrix. 
This expectation is confirmed for $t_p/t_l = 1/1000$, as the $\phi$-spectrum and the probability density of $\pi$-corner modes plotted in Fig.~\ref{fig:non_ideal_leads}(a) resemble the one in Fig.~\ref{fig:3DHOTI_r}(a) calculated for $t_p = 0$. 
Larger values of $t_p/t_l$ cause spreading of all reflection matrix eigenstates in space, and this leads to a range of effects.
  
The gap at $\phi = 0$ is closed first, signifying a topological phase transition that is followed by the appearance of four $0$-modes pinned to the corners of the system. 
This is shown in Figs.~\ref{fig:non_ideal_leads}(c) and (d). 
With larger values of $t_p$, the localization length of these $0$-modes increase, resulting in splitting of their eigenphases away from $\phi = 0$, see Fig.~\ref{fig:non_ideal_leads}(e), and resulting in a trivial phase, as clear from Fig.~\ref{fig:non_ideal_leads}(f).

Lastly, we discuss  the effect of the transverse coupling on the $\pi$-modes. 
In the limit $t_p = 0$, these modes had a localization length $\xi_0$ that was determined by the properties of the scattering region. 
The presence of transverse coupling causes the increase of this localization length,
also visible in the insets of Figs.~\ref{fig:non_ideal_leads}(a-c).  
As $t_p$ increases further, the large localization length causes the eigenphases corresponding to these corner states to split away from $\phi = \pi$, as seen from Fig.~\ref{fig:non_ideal_leads}(d), eventually leading to a trivial phase [Fig.~\ref{fig:non_ideal_leads}(e)].

Above, we have chosen a transversal coupling $t_p$ which is added to the first 10 sites of the lead. 
Its effect is to cause the propagating modes initially located on a single waveguide to spread in the \textit{y} and \textit{z} direction.
When this spread is comparable to the system size, the reflection matrix spectrum becomes trivial, since an incoming mode no longer probes the scattering region locally, e.g. at a single corner. 
However, as long as incoming modes, or linear superpositions thereof, are local in the y and z directions, we expect the topological features of the reflection matrix to remain intact.
This is because, for instance, the presence of zero-energy corner states in the 3D HOTP must be associated to $\pi$-modes in the reflection matrix due to resonant scattering.

\section{Nested scattering matrix formula} \label{app:Floquet_smatrix}

The Eq.~\eqref{eq:new_scat_matrix} is a way of determining the scattering matrix of a periodically-driven system by starting from its Floquet operator, ${\cal F}$.  
In our case, we use the same formula and replace the Floquet operator with the reflection matrix, $r_{\text{2D}}$, since we are interested in the analogy between the two. For simplicity and to connect with the physics of time-periodic phases, we will use ${ \cal F}$ throughout this appendix.

In the language of periodically driven systems, the Floquet operator ${ \cal F}$ has the effect of time-evolving a given state by one period of the drive, so from time $t$ to $t+T$. 
The projection operator $P$ serves to remove any portion of the state which overlaps with the absorbing terminals. 
In Eq.~\eqref{eq:new_scat_matrix} the matrix inverse is obtained as an infinite sum, $(\mathbbm{1}-X)^{-1}=\mathbbm{1}+X+X^2+X^3+\ldots$, a geometric series of terms involving time-evolution and projection operators. 
Physically, Eq.~\eqref{eq:new_scat_matrix} results from the fact that the absorbing terminals act stroboscopically on the wavefunctions of the driven system, that is, they are only active at discrete times, $t=nT$, with integer $n$.

Focusing on the $\phi=0$ case for simplicity, the first term in the geometric series reads $P {\cal F} P^ T$. 
Thus, an initial ($t=0$) state located on the sites corresponding to the absorbing terminals is time-evolved for one period, by applying ${\cal F}$. 
That part of the state which is still located on the sites of the absorbing terminals at $t=T$ is then projected out ($P$). 
The rest of the state, meaning that portion which was not located on the absorbing sites, $\mathbbm{1} - P^TP$, is again evolved for one period (up to $t=2T$), after which the projector onto the terminals is applied. 
This leads to the second term of the series, $P {\cal F} (\mathbbm{1} - P^TP) {\cal F} P^T$. 
The third term in the series again repeats this process. 
The leftover part of the state is again time-evolved and projected out if it overlaps with the absorbing terminals, leading to $P {\cal F} (\mathbbm{1} - P^TP) {\cal F} (\mathbbm{1} - P^TP) {\cal F} P^T$. 
Summing over all of these processes up to infinitely many driving periods leads to the geometric series whose sum is the inverse matrix appearing in Eq.~\eqref{eq:new_scat_matrix}. 
The resulting scattering matrix is unitary, as can be checked directly. 
This implies that, when summing up to infinitely many periods, all of the initial state is eventually projected out.

\bibliography{References}
\end{document}